\documentclass{jmoarticle}

\jmoTitle{Mechanistic determination of tear film thinning via fitting simplified models to tear breakup}
\jmoRunningTitle{Fitting simplified models to tear breakup}

\jmoAddAffiliation{1}{Department of Mathematical Sciences, University of Delaware, Newark, USA}
\jmoAddAffiliation{2}{School of Optometry, Indiana University, Bloomington, USA}

\jmoAddAuthor{1}{Rayanne A.}{Luke}
\jmoAddAuthor{1}{Richard J.}{Braun}
\jmoAddAuthor{2}{Carolyn G.}{Begley}

\jmoCorrespondence{Department of Mathematical Sciences, University of Delaware,
222 S. Chapel St, Newark, USA}

\jmoCorrespondenceEmail{rayanne@udel.edu}

\jmoAddKeyword{dry eye}
\jmoAddKeyword{fluorescent imaging}
\jmoAddKeyword{optimization}
\jmoAddKeyword{tear breakup}
\jmoAddKeyword{tear film}

\jmoAbstractPurpose{To determine whether evaporation, tangential flow, or a combination of the two cause tear film breakup in a variety of instances; to estimate related breakup parameters that cannot be measured in breakup during subject trials; and to validate our procedure against previous work.}
\jmoAbstractMethods{Five ordinary differential equation models for tear film thinning were designed that model evaporation, osmosis, and various types of flow. Eight tear film breakup instances of five healthy subjects that were identified in fluorescence images in previous work were fit with these five models. The fitting procedure used a nonlinear least squares optimization that minimized the difference of the computed theoretical fluorescent intensity from the models and the experimental fluorescent intensity from the images. The optimization was conducted over the evaporation rate and up to three flow rate parameters. The smallest norm of the difference was determined to correspond to the model that best explained the tear film dynamics.}
\jmoAbstractResults{All of the breakup instances were best fit by models with time-dependent flow. Our optimal parameter values and thinning rate and fluid flow profiles compare well with previous partial differential equation model results in most instances.}
\jmoAbstractConclusion{Our fitting procedure suggests that the combination of the Marangoni effect and evaporation cause most of the breakup instances. Comparison with results from previous work suggests that the simplified models can capture the essential tear film dynamics in most cases, thereby validating this procedure as one that could be used on many other instances.}

\lstdefinestyle{latex}{
  language=[LaTeX]TeX,
  escapeinside={\%*}{*)},
}
\lstdefinelanguage{bibtex}
{keywords={%
        @article,@book,@collectedbook,@conference,@electronic,@ieeetranbstctl,%
        @inbook,@incollectedbook,@incollection,@injournal,@inproceedings,%
        @manual,@mastersthesis,@misc,@patent,@periodical,@phdthesis,@preamble,%
        @proceedings,@standard,@string,@techreport,@unpublished%
    },
    comment=[l][shape]{@comment},
    sensitive=false,
}

\lstnewenvironment{lstlistingtex}{\lstset{language=[LaTeX]TeX}}{}
\usepackage{subfig}
\DeclareMathOperator*{\argmin}{argmin}
\usepackage{amsmath}
\usepackage{amssymb}
\usepackage{float}
\usepackage{expl3}
\expandafter\def\csname ver@l3regex.sty\endcsname{}

\begin{document}

\jmoMakeFrontpage



\section{Introduction}
\label{sec:intro}

Tear film breakup (TBU) occurs when a thinned region forms in the tear film (TF).  Clinically, this is defined as the first
dark area that is observed in the fluorescent (FL) TF following instillation of fluoresecein dye\cite{norn1969}.
Various mechanisms are thought to cause different types of TBU: evaporation\cite{lemp2007,willcox2017tf,king2018} causes relatively slow thinning\cite{king2010}, and rapid thinning may be explained by Marangoni-driven tangential flow\cite{zhong2019} or, plausibly, dewetting in cases of circular TBU\cite{yokoi2013,yokoi2019}.
The Marangoni effect drives outward flow at the aqueous/lipid interface induced by shear stress as a 
result of a lipid concentration gradient\cite{berger74,craster2009}.  Dewetting from a defective corneal 
surface region has been hypothesized
to drive outward tangential flow from pressure gradients due to van der Waals type forces \cite{sharma85,sharma99,zhang03}.
A related term is full-thickness tear film breakup (FT-TBU), which is when the aqueous layer has thinned to the point where the lipid layer and glycocalyx touch\cite{begley13,king2018}. 

The effects of evaporation and the Marangoni effect on TBU have been extensively studied and modeled separately\cite{king2013,peng2014,braun2018,zhong2018}; only recently have they been explored in combination to explain breakup occurring on an intermediate time scale\cite{zhong2019}.  Zhong \textit{et al.}\cite{zhong2019} developed a partial differential equation (PDE) model with one spatial variable that incorporated both mechanisms.
Luke \textit{et al.}\cite{luke2021} fit FT-TBU data from fluorescent (FL) images from healthy subjects with a rescaled version of the Zhong \textit{et al.}\cite{zhong2019} model. The optimization was conducted via nonlinear least squares minimization of the theoretical and experimental FL intensity, and TBU parameters were estimated for the model. The PDE fitting process is time-consuming and limited to spots or streaks; more complicated shapes could be fit using two spatial dimensions.

Ordinary differential equation (ODE) models have been designed to capture TF thinning without spatial variation; exact solutions exist for some cases. Braun \textit{et al.}\cite{braun2019} extended an ODE model without flow from previous work\cite{braun2014} to include time-independent extensional fluid flow; a time-dependent version has recently been developed and is presented here. The flow is divergent from the origin and can be considered with evaporative loss and osmotic supply. Winter \emph{et al.}\cite{winter2010} (PDE model) and Braun\cite{braun2012} (ODE model) included van der Waals forces to stop thinning with a zero permeability condition at the tear/cornea interface; such terms and forces are omitted from the models in this work.  Luke \textit{et al.}\cite{luke2020} fit TBU data with ODE models with evaporation, with or without osmosis, but without tangential flow. Neither the PDE nor ODE model gave the best fit for all of the TBU instances. The instances best fit by the ODE models had rates of FL intensity decrease most closely approximated by a constant.

Both TF instability and hyperosmolarity are important to study because they are proposed as etiological causes of dry eye syndrome \cite{gilbard1978,craig2017defn, willcox2017tf}. Osmolarity is the osmotically-active salt ion concentration in the aqueous layer\cite{tomlinson09,stahl2012}. A concentration difference between the corneal epithelium and aqueous layer induces osmotic flow from the cornea to the TF\cite{peng2014,braun2015}. TF osmolarity may be measured in the inferior meniscus clinically\cite{lemp2011}; the healthy range is 296-302 Osm/(m$^3)$\cite{lemp2011,tomlinson2006,versura2010}. Dry eye measurements in the same location can reach 316-360 mOsm/(m$^3)$\cite{gilbard1978,tomlinson2006,sullivan2010,dartt2013} but estimates for the TF over the cornea reach 900 mOsm/m$^3$ or higher\cite{liu09,braun2015,peng2014,luke2020}. High levels of TF osmolarity are associated with pain, inflammation and cellular changes\cite{pflugfelder2011,belmonte2015,liu09}. In support of these potentially high levels of TF osmolarity over the cornea, mathematical models without spatial variation have estimated peak osmolarities up to ten times the isotonic concentration\cite{braun2012,braun2015}. The modeling work of Peng \textit{et al.}\cite{peng2014} found that evaporation elevates osmolarity in breakup regions.

TF thinning rates have been measured experimentally or estimated in many studies. A few experimental methods include spectral interferometry \cite{nichols2005,kimball2010,king2010}, an open chamber\cite{hamano1981}, an evaporimeter\cite{peng2014b}, and averaging pure water and functioning lipid layer rates over the cornea obtained by heat transfer analysis and thermal imaging\cite{dursch2018}.
In Braun \textit{et al.}\cite{braun2018}, both peak and background evaporation rates in TBU instances, as well as the width of the evaporation distribution, were studied parametrically.  Subsequently, parameter estimation schemes were developed in Luke \textit{et al.}\cite{luke2020,luke2021} for fitting PDE models to experimental FL intensity distributions.  They found evaporation rates ranging from -36.9 to 4.91 $\mu$m/min (the upper bound indicating thickening) and overall TF thinning rates ranging from -23.5 to -1.85 $\mu$m/min. These thinning rates were comparable to, or a little faster than, previous experimental rates measured there were not specifically seeking TBU instances \cite{nichols2005}.

 In this paper, we fit a hierarchy of ODE models to the same dataset as in Luke \textit{et al.}\cite{luke2021}. The authors fit TBU instances with PDE models that incorporated evaporation and the Marangoni effect\cite{luke2021}. We use these PDE results as a guide when determining whether our results have captured what we believe to be the correct dynamics. In some cases, the ODE models are better able to follow the experimental data than the PDEs, suggesting different conclusions may be drawn for those particular instances.

\section{Methods}
\label{sec:methods}

\subsection{FL images}
\label{sec:images}

The data was taken from twenty-five normal, healthy subjects in a study conducted at Indiana University \cite{awisigyau2020} as discussed in several papers\cite{luke2020,luke2021}. Approval was received from the Biomedical Institutional Review Board of Indiana University, Declaration of Helsinki principles were followed during data collection, and informed consent was obtained from all subjects. Subjects were seated behind a slit lamp biomicroscope and 2\% sodium fluorescein solution was instilled in the patient's eye. A light with a cobalt blue excitation filter illuminated the eye so that the aqueous layer of the TF fluoresced green\cite{carlson2004}. A trial is the sequence of images of the subject's eye following a few quick blinks. The trial records the fluorescence of the aqueous part of the TF. The trials typically start with an FL concentration close to 0.2\%, which is the so-called critical concentration where peak fluorescence occurs for thin TFs\cite{webber86}. The critical FL concentration can be expressed in molar as 0.0053 M; see Appendix \ref{sec:scale}.

\begin{figure}
\centering
\subfloat[][S9v1t4]{\includegraphics[scale=.08]{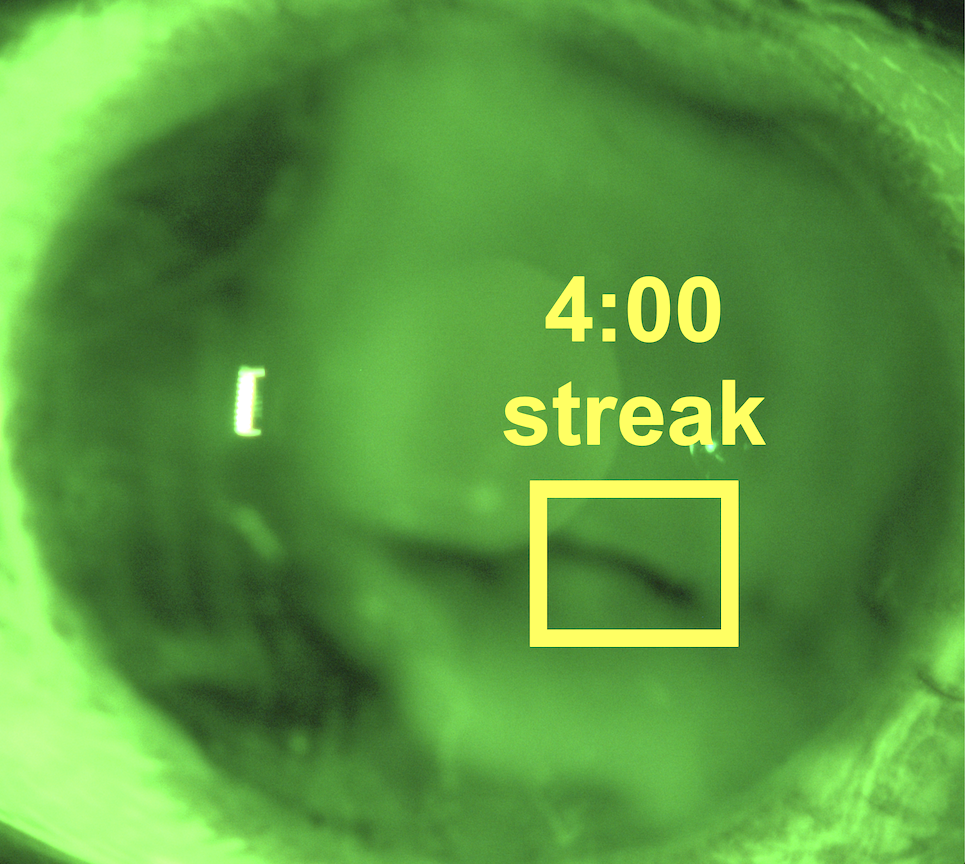}}
\subfloat[][S9v2t1]{\includegraphics[scale=.08]{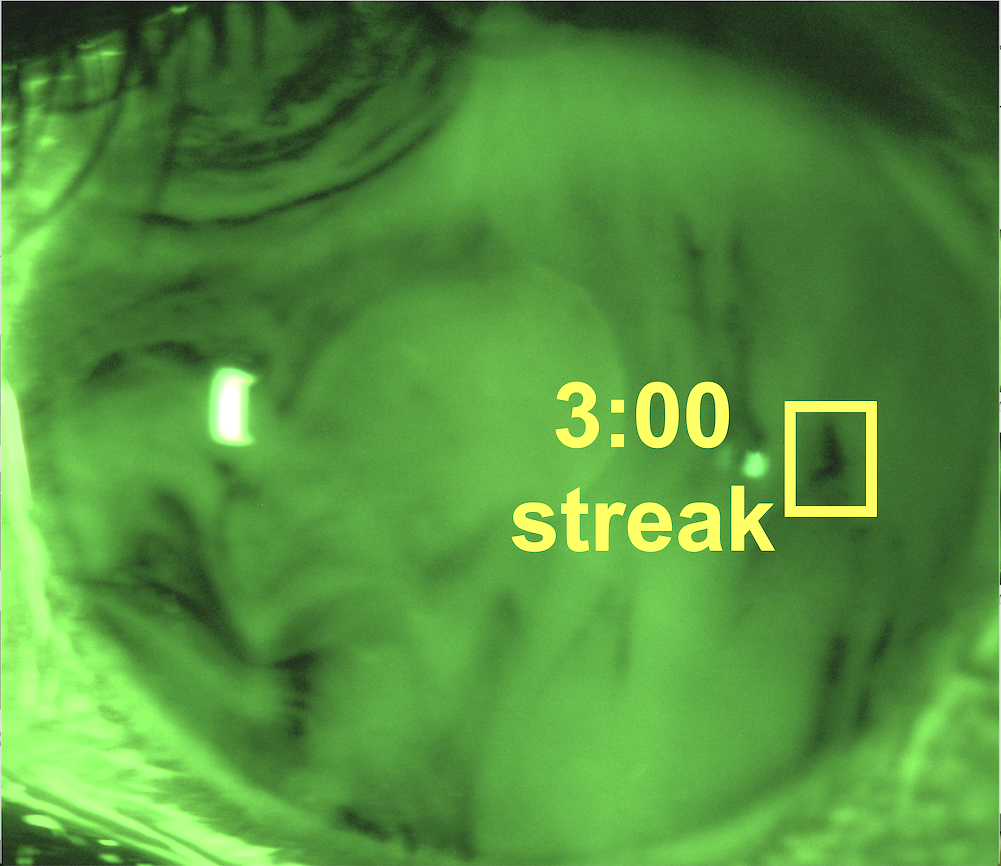}}
\subfloat[][S9v2t5]{\includegraphics[scale=.08]{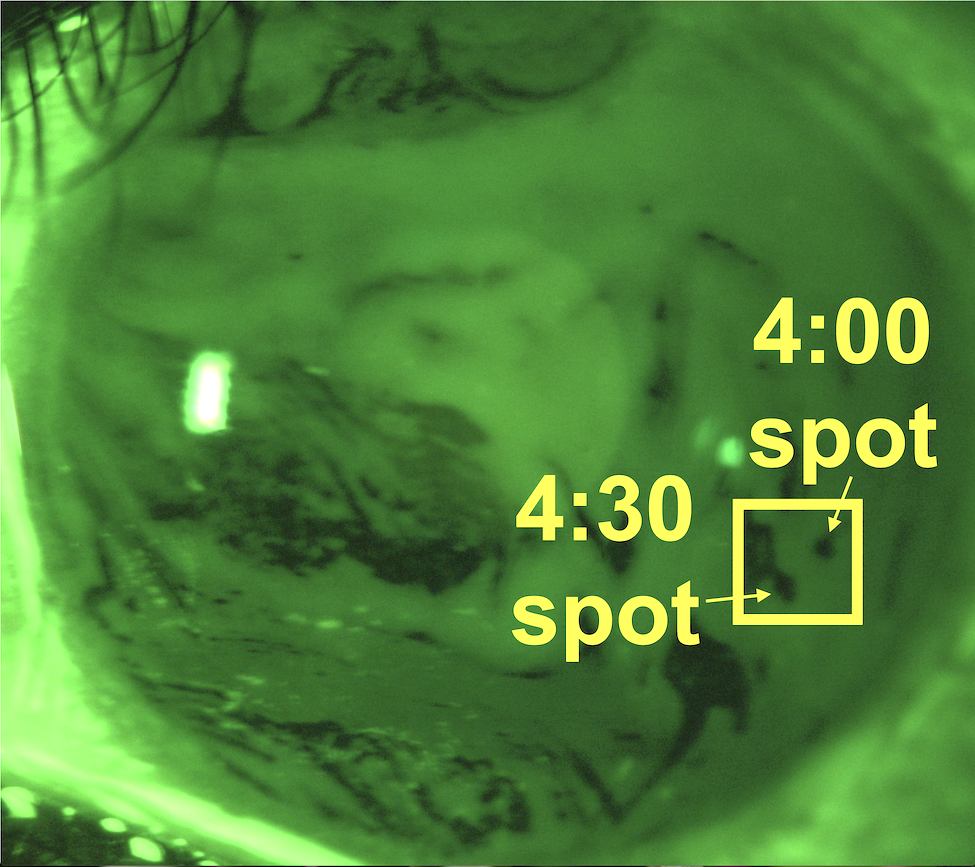}} \\
\subfloat[][S10v1t6]{\includegraphics[scale=.08]{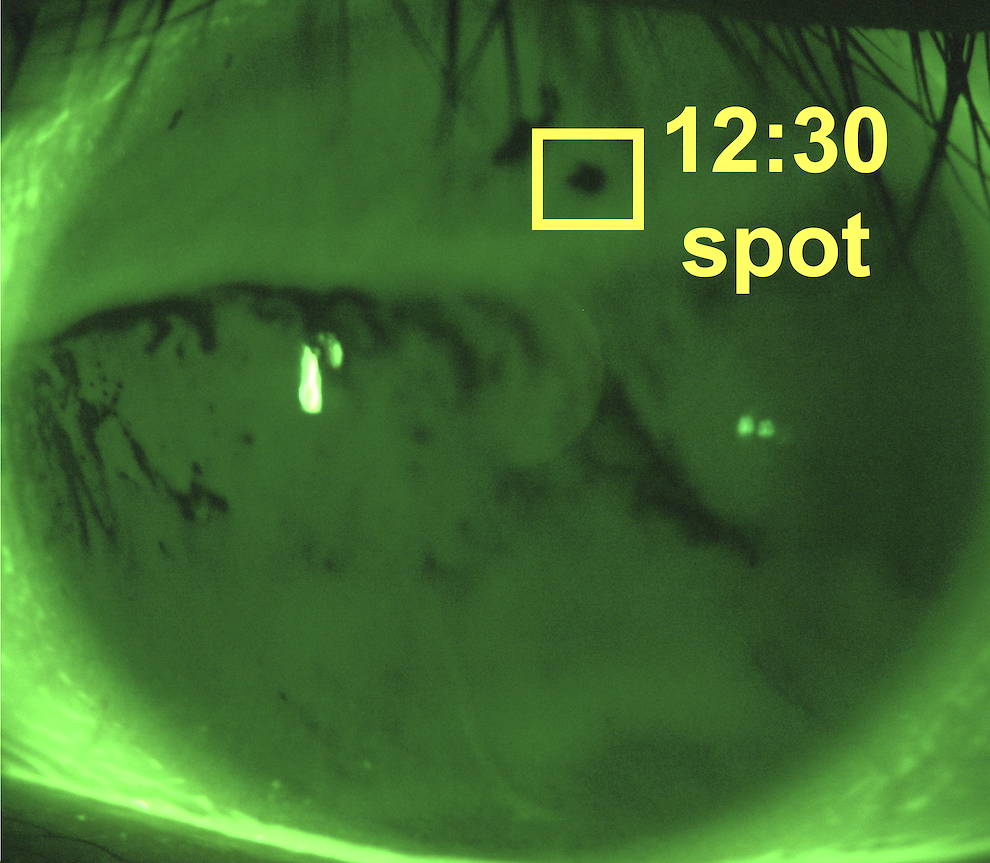}}
\subfloat[][S13v2t10]{\includegraphics[scale=.08]{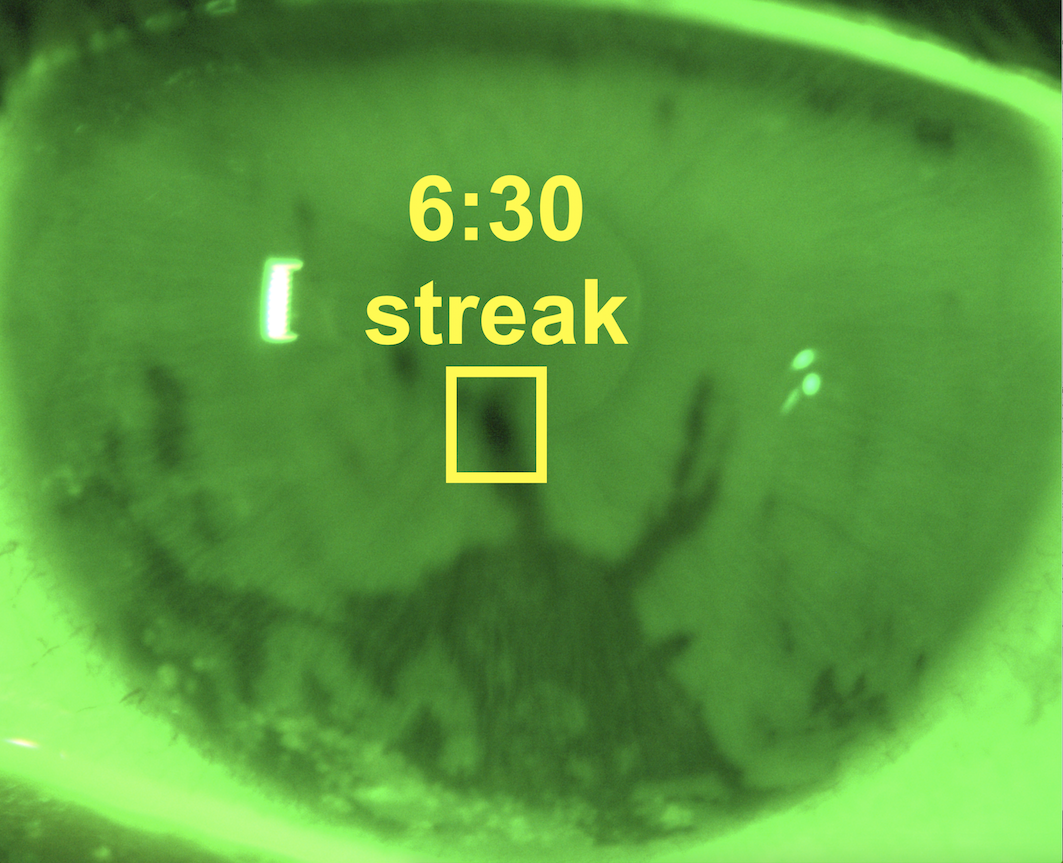}}
\subfloat[][S18v2t4]{\includegraphics[scale=.08]{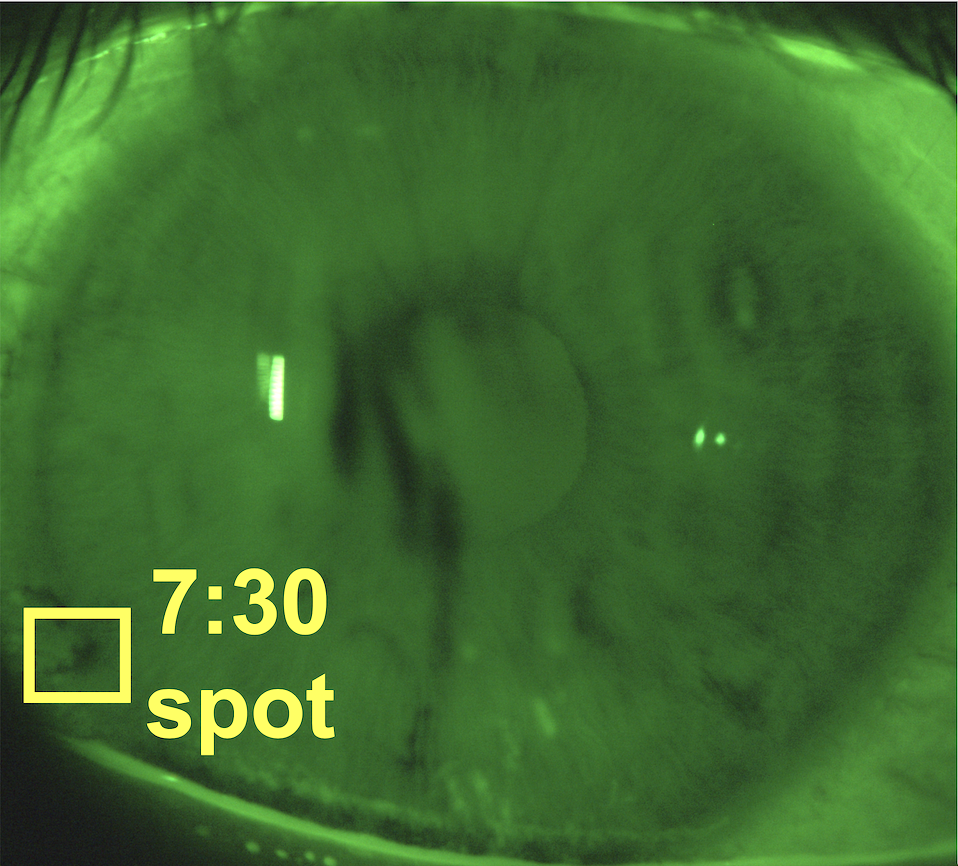}} \\
\subfloat[][S27v2t2]{\includegraphics[scale=.17]{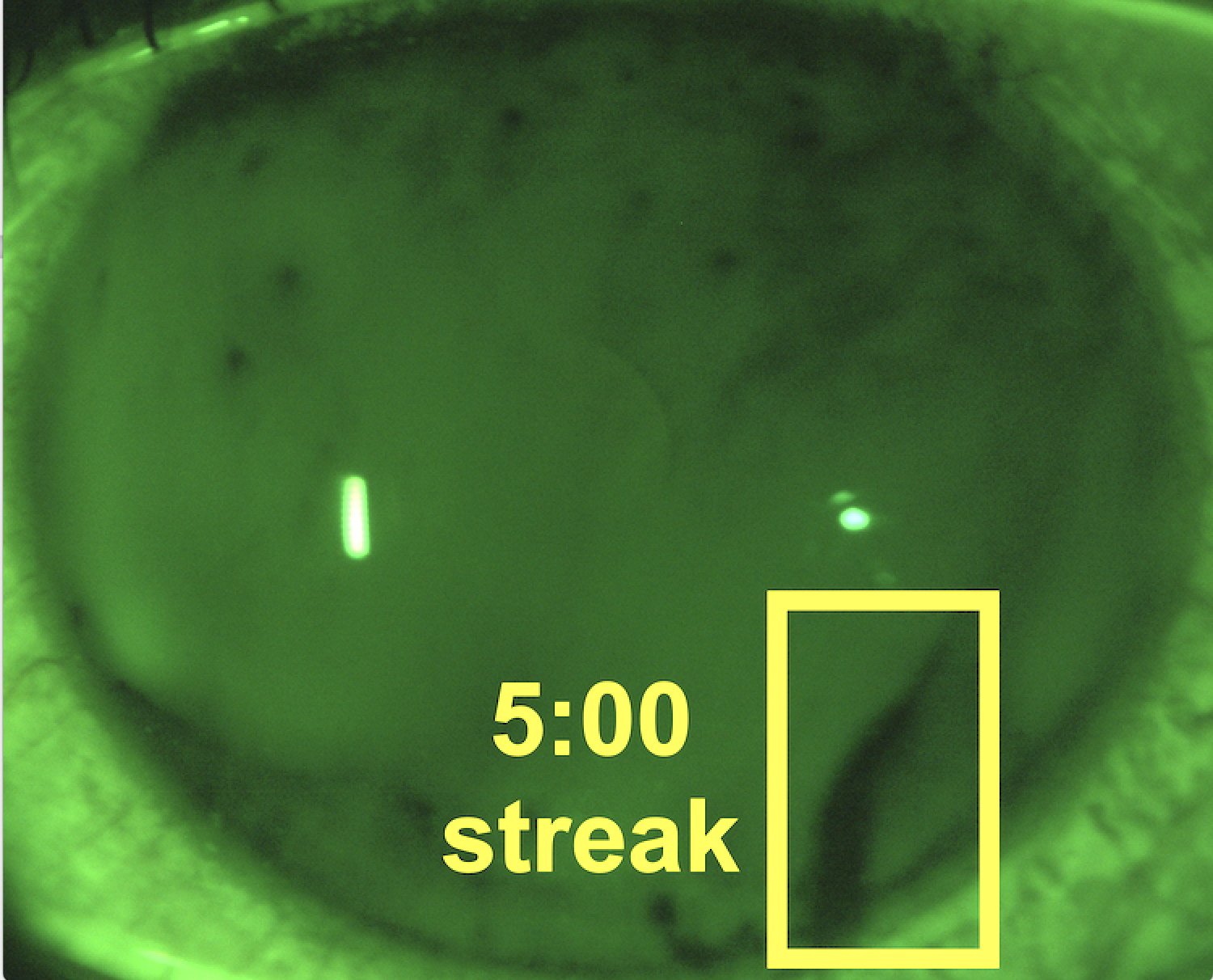}}
\subfloat[][S27v2t2 5:00 streak surface plot]{\includegraphics[scale=.128]{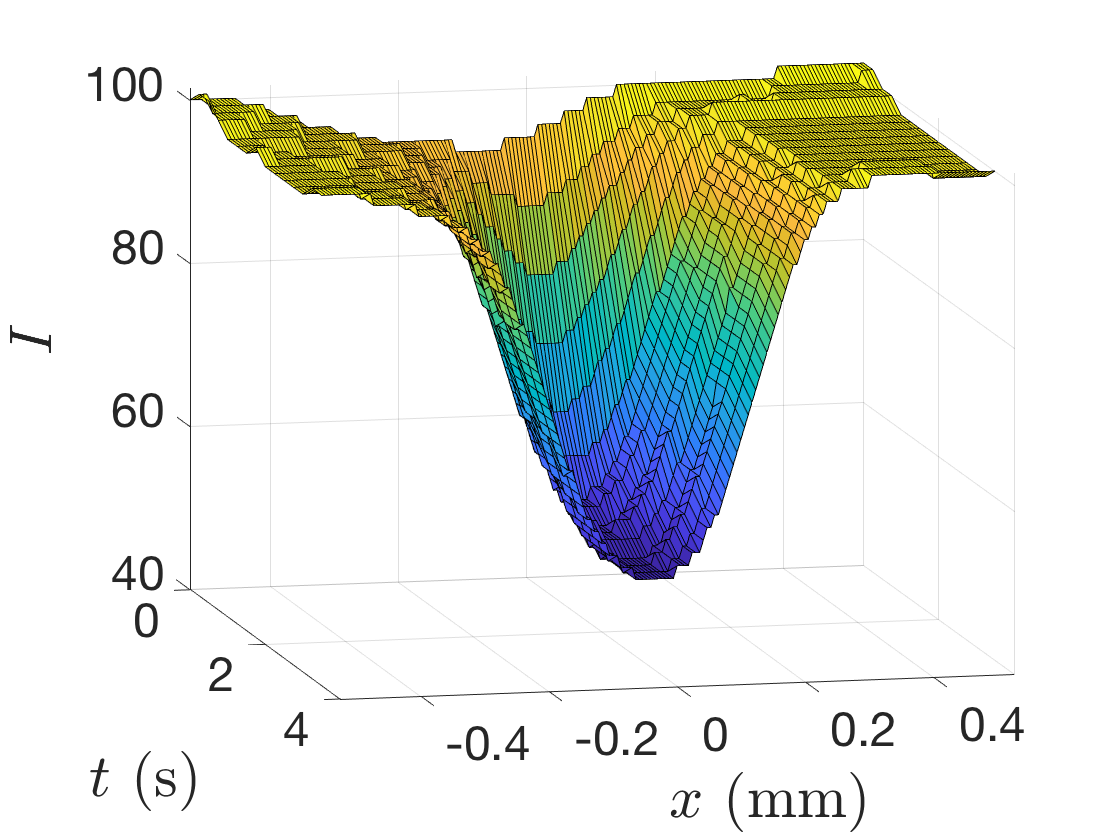}}
\caption{\footnotesize{(a)-(g): the last image in each trial. The bright rectangle, called the Purkinje image, is due to the reflection from the light source. The images have been brightened for better visibility. (h): Surface plot of the FL intensity over time for subject S27v2t2 5:00 shown in (g).}}
\label{fig:images}
\end{figure}

For our purposes, FT-TBU is thinning to what is evidently a very small aqueous thickness, as determined by the aid of a computer. We fit the central data of the same extracted data from spot- or streak-shaped FT-TBU instances in Luke \textit{et al.}\cite{luke2021}. All instances reported in this paper are shown in Figure \ref{fig:images}.
The time resolution of our dynamic data is restricted by the frame rate, which is 4 or 5 $s^{-1}$ depending on the trial.

\subsection{Models}
\label{sec:model}

There is a hierarchy of ODE models we explore; the most complicated is derived in Appendix \ref{sec:app_deriv}. Each variation of the model is determined by which mechanisms are assumed to affect the TF. The options are a combination of evaporation, osmosis, and flow. If flow is present, there are several choices we use. These simple models are designed to capture the key ingredients in thinning in order to distinguish which is the dominant mechanism causing thinning: evaporation, outward tangential flow, or a combination of the two. Evaporation-dominated thinning is characterized by inward tangential flow, if any\cite{luke2020}, while Marangoni flow is characterized by strong outward tangential flow that decreases in strength as the trial progresses\cite{luke2021}.   In all models that follow, $h(t)$ denotes the nondimensional TF thickness.

\subsubsection{Initial conditions}

In all cases of the model, we have uniform nondimensional initial conditions
\begin{equation}
c(0) = 1,\  h(0) = 1,  \mbox{ and } f(0) = f_0.
\end{equation}
After eliminating $c$ and $f$, the only initial condition needed is the one for $h$.

\subsubsection{Case E model}

The simplest variations of the model assumes constant evaporation is the only mechanism affecting the TF thickness. The nondimensional evaporation rate is $v$. The differential equation for $h$ , which is conservation of water, is given by
\begin{equation}
\dot{h}= - v.
\end{equation}
A dot indicates a time derivative.

\subsubsection{Case O model}

We assume constant evaporation and osmosis affect the TF thickness. Figure \ref{fig:sketch} shows a sketch of the model (on left).  Osmolarity is quantified in the nondimensional variable $c$. Osmosis is written as a concentration difference between the cornea and the TF; a greater TF osmolarity will drive osmotic flow from the cornea into the TF. The differential equation is given by
\begin{equation}
\dot{h} = - v + P_c (c-1),
\end{equation}
where $P_c$ is nondimensional permeability constant.  Mass conservation of solute (osmolarity), namely, $ch = 1$, allows us to eliminate $c$\cite{braun2014} to obtain a single ODE for $h$:
\begin{equation}
\dot{h} = - v + P_c \left[ \frac{1}{h} - 1 \right].
\end{equation}
Cases E and O model situations where evaporation is the dominant mechanism affecting TF thinning and where flow is not important.

\subsubsection{Case F model} 
\begin{figure}
 \centering
 \includegraphics[scale=.25]{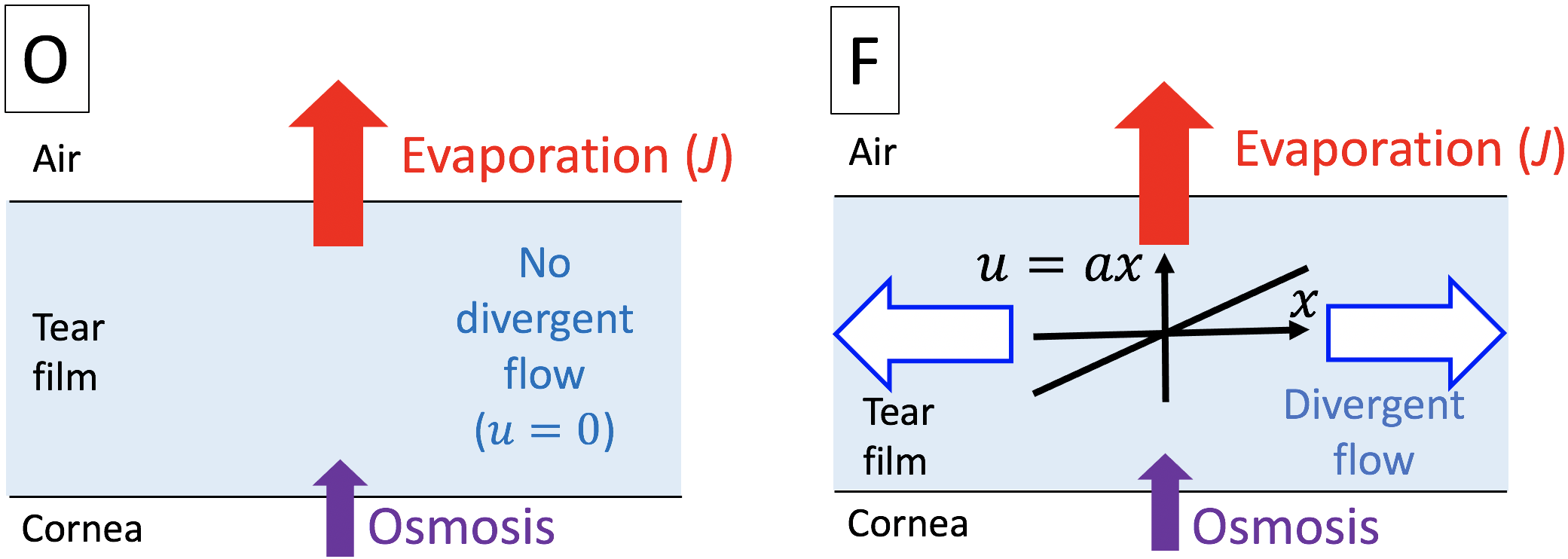}
 \caption{Schematic for the Case O and F models.}
 \label{fig:sketch}
 \end{figure}

We assume that evaporation, osmosis, and flow affect the thickness of the TF. Figure \ref{fig:sketch} shows a sketch of this model (on right).  We introduce nondimensional velocity along the TF, $u(x,t)$. In this first case,
\begin{equation}
u(x,t) = u(x) = ax,
\label{eq:flow_f}
\end{equation}
where $\partial_x u = a$ is the strain rate. This flow may be thought of as stretching taffy if $a > 0$. A single curve of Figure \ref{fig:flow} illustrates this flow profile. Mass conservation becomes $ch = e^{-at}$, and the differential equation for the TF thickness is
\begin{equation}
\dot{h} = -v + P_c \left[ \frac{\exp(-at)}{h} - 1 \right] - ah.
\label{eq:dhdt_f}
\end{equation}
 For this and all following models, solute conservation relates the derivative of the solute mass ($hc$) to the strain rate. For the case F model, this is given by
\begin{equation}
\dot{h} = - ( \partial_x u )  h = -ah.
\end{equation}
 The sign of the strain rate $a$ suggests the kind of flow present. If $a < 0$, the flow is inward, mimicking healing capillary flow. This characterizes evaporation-dominated thinning. If $a > 0$, the flow is outward, mimicking Marangoni flow, driven by interfacial shear stress. This characterizes Marangoni effect-dominated thinning.

 \subsubsection{Case D model} This model is designed to mimic the time-dependent flow seen in TBU instances where flow is dominated by the Marangoni effect. We assume evaporation, osmosis, and decaying extensional flow affect the TF thickness. Here,
 \begin{equation}
 u(x,t) = b_1  e^{-b_2 t}x,
 \label{eq:flow_d}
 \end{equation}
 where $b_1$ and $b_2$ are flow rate and decay rate parameters, respectively. In this case, the strain rate is $ \partial_x u = b_1 e^{-b_2 t}$. The differential equation for TF thickness $h$ is
 \begin{equation}
 \dot{h} =  - v + P_c\left\{\frac{1}{h} \exp \left[ \frac{b_1}{b_2}  \left( e^{-b_2 t} - 1 \right) \right] -1 \right\}  -b_1 e^{-b_2t} h,
 \end{equation}
 where the first term inside the brackets is the result of using mass conservation to eliminate $c$.

 \subsubsection{Case M model} 
\begin{figure}
\centering
\includegraphics[scale=.2]{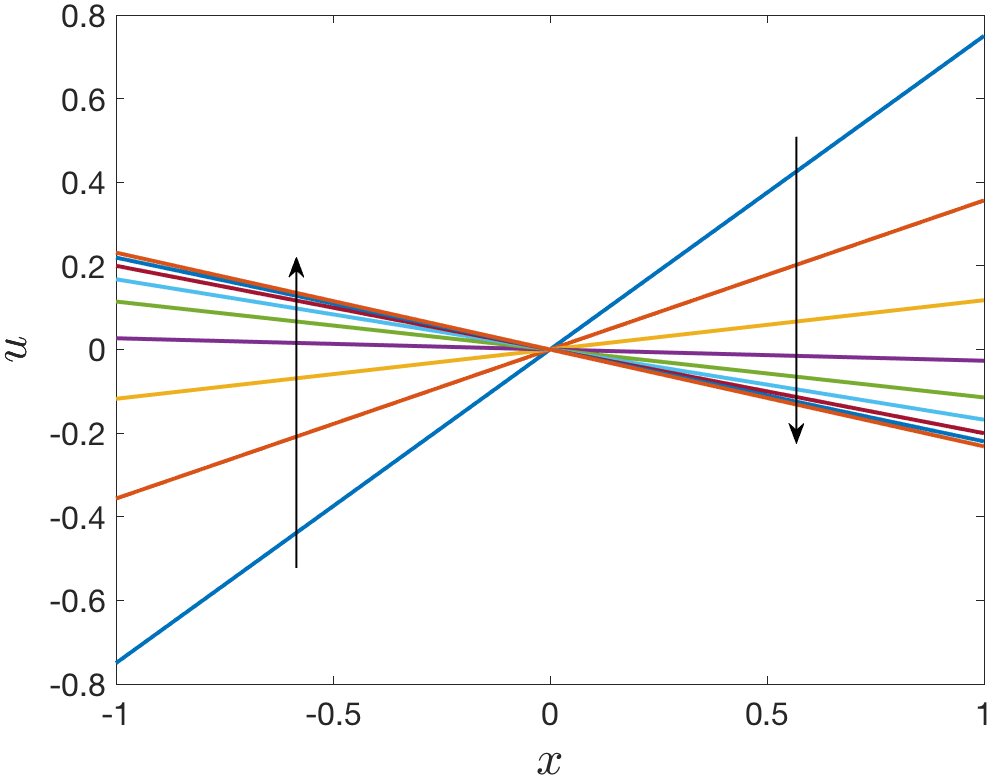}
\caption{\footnotesize{An example of the fluid flow profile $u$ from Equation \ref{eq:flow}. This simulation models strong outward tangential flow that dies off and then inward capillary flow persists. The nondimensional parameters used are $a = -0.25, b_1 = 1,$ and $b_2 = 2$. A final nondimensional time of 2 was used. Arrows indicate increasing time.}}
\label{fig:flow}
\end{figure}

Our most complicated the model is an extension of case D to allow the flow to decay to a constant, nonzero value. We assume that evaporation, osmosis, and a combination of constant and decaying extensional flow affect the TF thickness. This model allows for the flow to change direction: for example, it may start outward and strong, but then decay to a weakly inward, constant value. The nondimensional horizontal fluid profile is given by

\begin{equation}
u(x,t) = (a + b_1e^{-b_2 t})x.
\label{eq:flow}
\end{equation}
The exponential term will greatly diminish that part of the flow after $1/b_2$ units of time. The strain rate is $\partial_x u = a + b_1 e^{-b_2 t}$. An example of this fluid profile is shown in Figure \ref{fig:flow}.
The nondimensional film thickness, $h(t)$, is governed by
\begin{equation}
\dot{h} = - v + P_c\left\{\frac{1}{h} \exp\left[ - a t + \frac{b_1}{b_2} \left(e^{-b_2 t} - 1 \right) \right] -1 \right\}  -(a + b_1 e^{-b_2 t}) h.
\label{eq:mod_5}
\end{equation}
Appendix \ref{sec:app_deriv} shows how $c$ can be eliminated in \ref{eq:mod_5} via solute conservation.

\subsection{FL intensity}

Nondimensional FL intensity $I$ from the TF is given by
\begin{equation}
I = I_0 \frac{1 - \exp(-\phi h f)}{(1 + f^2)}.
\label{eq:I}
\end{equation}
Similarly as for the equation for $h$, $f$ can be eliminated so that the FL intensity $I$ for the Case M model is given by
\begin{equation}
I = I_0 \frac{ 1 - \exp \left\{- \phi f_0 \exp \left[ - a t +  \frac{b_1}{b_2} \left( e^{-b_2 t} - 1 \right) \right] \right\}}{ 1 +  \left\{ f_0 \exp \left[ -at + \frac{b_1}{b_2} \left( e^{-b_2 t} - 1 \right) \right]/ h \right\}^2}.
\end{equation}
The expressions for $I$  for the other cases can be obtained by setting $b_1 = 0$ if there is no time-dependence in the flow, by setting $a = 0$ if there is only time-dependent flow, and by setting $a = b_1 = 0$ if there is no flow at all.

\subsection{Estimating initial physical quantities}

We estimate the initial FL concentration following Wu \textit{et al.} \cite{wu2015}. This value is assumed to be uniform throughout the TF. By inverting the dimensional version of Equation \ref{eq:I} (Equation \ref{Idim} in Appendix \ref{sec:app_deriv}) for $h'$, we obtain an initial TF thickness estimate: 
\begin{equation}
h_0' = - \frac{1}{\epsilon_f f_0'} \log \left\{ 1 - \frac{I}{I_0} \left[ 1 + \left( \frac{f_0'}{f_{\text{cr}}}\right)^2 \right] \right\}. 
\end{equation}
Model eye calculations\cite{wu2015} determine $I_0$ through a least squares problem. The relative initial FL intensity $I$ at the center of the TBU instance, where the minimum FL intensity in the corneal region of interest has been subtracted off, is used. More details about the procedure can be found in Luke \textit{et al.}\cite{luke2020}.

\subsection{Optimization}

We follow the process described in Luke \textit{et al.}\cite{luke2020}; a summary is given below.

\subsubsection{ Data preparation}

We use custom MATLAB codes to convert the images in each trial from RGB color to grayscale, smooth the images with a Gaussian filter, and stabilize the images using the Purkinje image\cite{awisigyau2020}, a bright artefact reflecting the light source. We use the same roughly linear or circular FT-TBU instances that were chosen and fit by a PDE model in Luke \textit{et al.}\cite{luke2021} to compare with the PDE results.
We fit our theoretical FL intensity to  the central data of a subset of about 6-10 time levels of experimental FL intensity data from the trial. The starting frame is the last frame before the FL intensity data starts decreasing. The first few frames of a trial are taken at a lower light setting to obtain an initial FL concentration estimate, and in some trials there is evidence that thinning has begun during this interval. As a result, the first bright image already exhibits significant decrease in FL intensity in the center of breakup. Luke \textit{et al.}\cite{luke2021} remedied this issue by introducing ``ghost'' time levels, allowing the model solution to start with a uniform time level that is not compared to the experimental FL data. This is a product of the low time resolution of our data. In this work, we also use ghost times as appropriate. The last frame is the final frame before the FL intensity data stop decreasing.

\subsubsection{ Optimization problem}

We discuss the optimization for the case M model. Expressed in continuous variables, we seek to minimize $||I_{\text{th}}(t)-I_{\text{ex}}(t)||_2^2$ over the parameters $v'$, the evaporation rate, $a'$, the constant extensional flow rate, $b_1'$, the decaying extensional flow rate, and $b_2'$, the decay rate.
Here, $t$ corresponds to the time after the light source brightness has been increased to the high setting. This variable has been nondimensionalized with the scalings given in Appendix \ref{sec:scale}. The norm is over all $t \in [0, T ]$ excluding any ``ghost'' time levels from the theoretical FL intensity, where $T$ corresponds to the total length of the trial.
The optimization problem may be written
\begin{equation}
\argmin_{v', a', b_1',b_2'} ||I_{\text{th}}(t;v',a', b_1',b_2')-I_{\text{ex}}(t)||_2^2.
\end{equation}
 Theoretical intensity, $I_{\text{th}}$, is computed after solving the differential equation for film thickness, $h$. Similar optimizations are conducted for each of the other models.

\subsubsection{ Numerical method and stopping criterion}

 The five ODEs for $h$ are solved using ode15s in MATLAB (MathWorks, Natick, MA, USA). For the optimization, we use a nonlinear least squares minimization implemented by lsqnonlin in MATLAB with the trust-region reflective algorithm\cite{nocedal2006} and we add a second order finite difference approximation of the Jacobian\cite{leveque2007} to improve performance.
To generate initial guesses for optimization, forward computations were conducted until the theoretical dynamics were close to the experimental. For each instance, the solver stopped because the change in residual was less than the specified tolerance. Optimization tolerances of roughly the square root of the solver tolerances were used.

\section{Results}
\label{sec:results}

\subsection{Exact solutions}

Exact solutions exist for the case E model and for case F with $P_c=0$. For the case E model, using our initial condition, the nondimensional exact solution is
\begin{equation}
h(t) = 1 - vt.
\end{equation}
This solution ignores the physical reality that evaporation ceases once the TF reaches zero thickness; thus, the solution is only relevant for $t \in [0, 1/v].$

If we assume that time-independent flow is the only mechanism affecting the TF thickness in case F, then $v = P_c=0$  in (6) applies:
\begin{equation}
\dot{h} = -ah.
\end{equation}
Using our initial condition, we find that
\begin{equation}
h(t) = e^{-at}.
\end{equation}
To model TF thinning, we assume $a > 0$ in this instance. Thus, as expected, the TF thins to zero as time increases\cite{howell1996}.

\subsection{Numerical solutions}

\begin{figure}
\centering
\subfloat[][TF thickness]{\includegraphics[scale=.15]{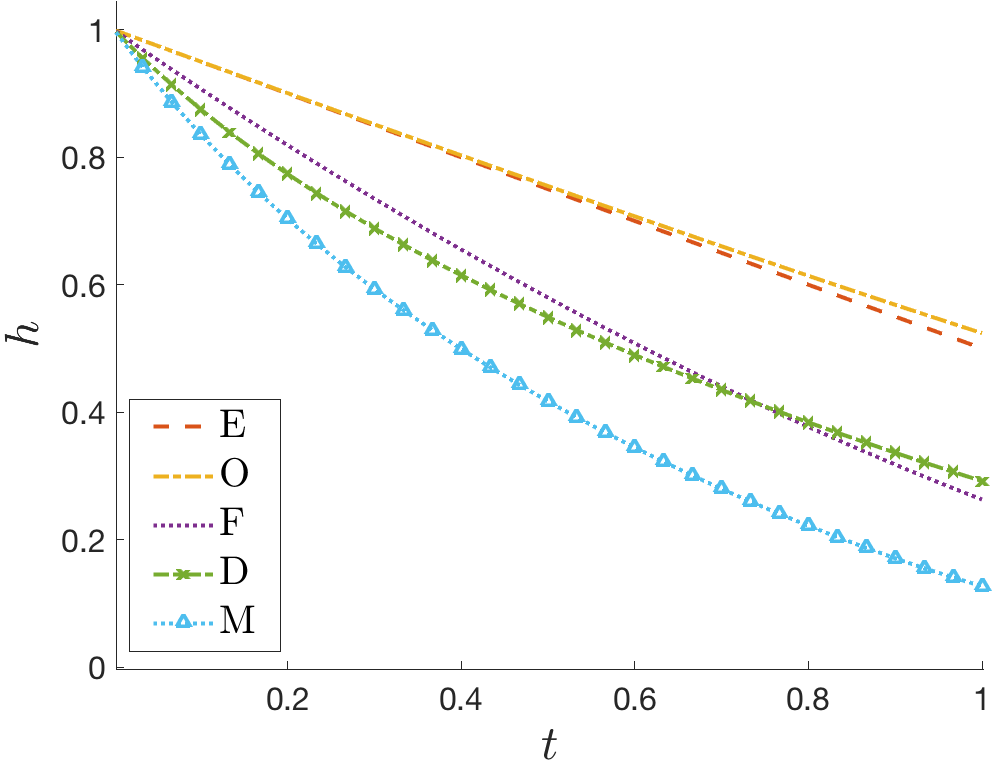}}
\subfloat[][FL intensity]{\includegraphics[scale=.15]{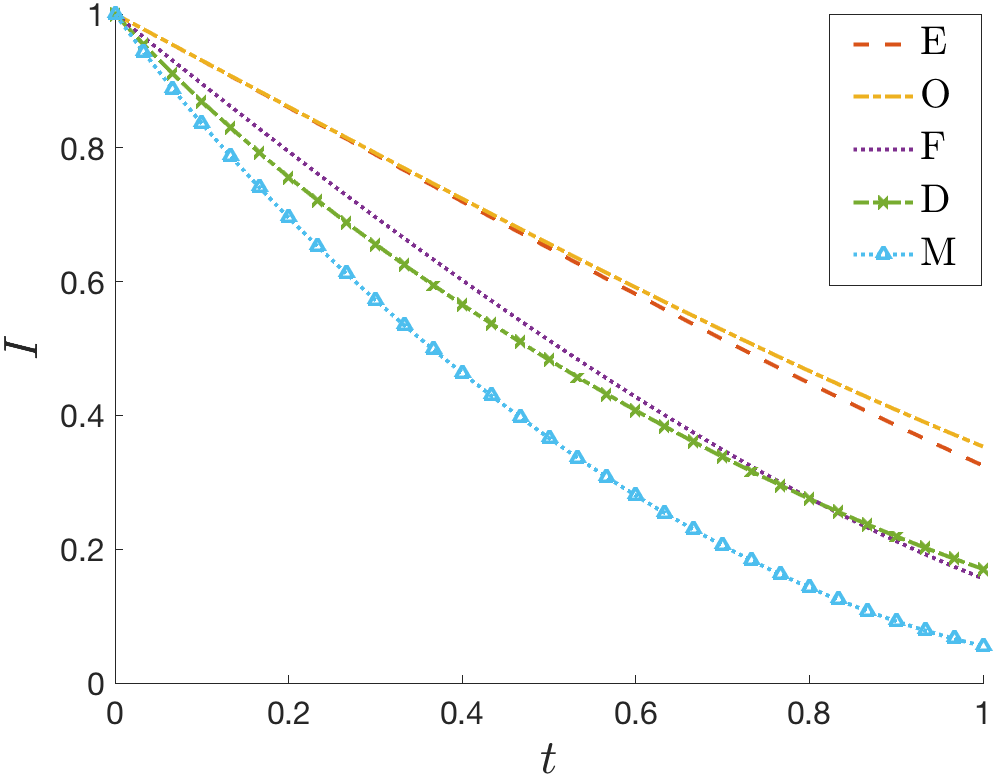}} \\
\subfloat[][Osmolarity]{\includegraphics[scale=.15]{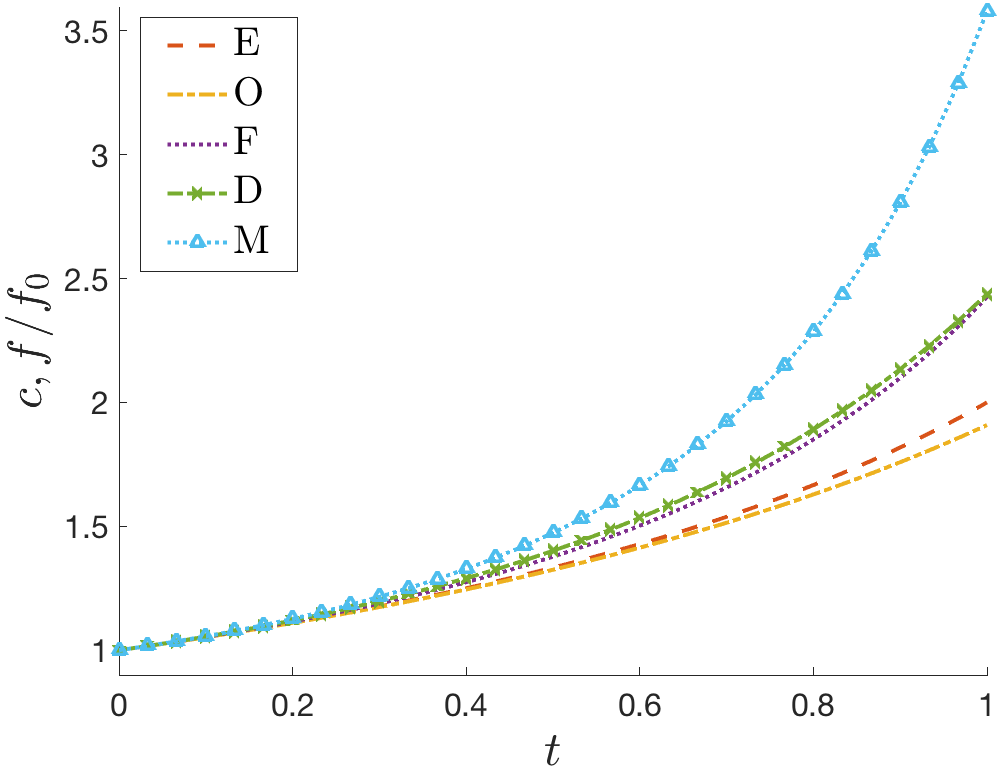}}
\caption{Nondimensional theoretical solutions for the five cases of the model with \\ $v' = 30 \ \mu$m/min, $a' = 0.15$ /s, $b_1' = 0.3$ /s, $b_2' = 0.8$ /s, $f_0' = 0.3$\%, $d = 3 \ \mu$m, and $t_s = 3$ s.}
\label{fig:plot_h}
\end{figure}

In Figure \ref{fig:plot_h} we plot nondimensional theoretical solutions for all of the models. For comparison purposes, we have used the same parameter values for each of the five models. In particular, both flow parameters are positive, indicating outward flow. The nondimensional parameters that result from our scalings are $a = 0.45, b_1 = 0.9, b_2 = 2.4, v = 0.5,$ and $P_c = 0.0653$. We see that the case O solution thins slightly less than the case E solution due to osmosis, which adds fluid to the TF. For the three models involving flow, since we have selected both $a, b_1 > 0$, the case M model shows the most thinning since the outward flow is strongest of all the models. As expected, the case F and D models are similar early on, but become increasingly similar once the flow is shut off in the D model. Osmolarity and normalized FL concentration solutions are identical in the absence of spatial variation. Both quantities are inversely related to TF thickness and increase at the origin in the presence of outward flow; the profiles reach the highest peak value for the case M model, which exhibits the greatest decrease in TF thickness and the strongest fluid flow.

\subsection{Fitting results}
\label{sec:fits}

We begin by presenting our aggregate results of fitting the same instances that were fit with the mixed-mechanism PDE model of Luke \textit{et al.}\cite{luke2021}. The best fit results as determined by the smallest norm are shown in Table \ref{table:results}. Each FT-TBU instance is labeled by subject, visit, and trial number, the location of the breakup as a clock reading, and the type of breakup (streak or spot). Images showing the FT-TBU instances can be found in Section \ref{sec:images}. A combination of the evaporation rate, constant flow rate, decaying flow rate, and decay rate are adjusted to accomplish the fit. The optimal parameters are given for the case of the model with the smallest norm. Section \ref{sec:examples} shows examples of the experimental data, fits, and resulting theoretical solutions using the optimal parameters found by nonlinear least squares minimization. The S18v2t4 7:30 spot was originally fit with a single ghost time level in Luke \textit{et al.}\cite{luke2021} but alternatively fit with two in the supplementary material; we choose to fit with  two here as well.

Five of the FT-TBU instances are best fit by the case M model and the other three are best fit by the case D model. Notably, the versions of the model without flow produce theoretical solutions with worse fits in all eight instances; we take this as strong evidence that flow plays a crucial role in causing the TF thinning. This is expected as each FT-TBU was previously fit with a model that combined evaporation and strong Marangoni flow\cite{luke2021}. It is worth mentioning a few other fits: the S27v2t2 5:00 streak is also fit well with the case D model, with the case F model not far behind; the case M model fits the S18v2t4 7:30 spot data well; and the S13v2t10 6:30 spot data also matches reasonably well with the case F model.

Table \ref{table:results} shows a wide range of evaporation rates.
Notably, the S9v2t5 4:30 spot instance, which was categorized as evaporation-dominated in Luke \textit{et al.}\cite{luke2021}, has the highest optimal evaporation rate. In contrast, the S10v1t6 12:30 spot and S27v2t2 5:00 streak are faster instances that were categorized as Marangoni effect-dominated in the aforementioned paper; these cases exhibit the two smallest evaporation rates seen in Table \ref{table:results}. The S18v2t4 7:30 and S10v1t6 12:30 spots have the strongest outward flow of all instances, with initial strain rates close to or over 2 $s^{-1}$. Further, all five Marangoni effect-dominated FT-TBUs from the previous paper exhibit outward flow, the characteristic direction of Marangoni flow. The S9v2t1 3:00 streak and S9v2t5 4:00 and 4:30 spots are the three instances designated as evaporation-dominated or transitional thinning in the PDE paper; these all show some amount of inward flow, which is characteristic of evaporation-dominated thinning.

\begin{table}[h]
  \centering
  \begin{tabularx}{\textwidth}{|p{.11\linewidth}|p{.08\linewidth}|p{.05\linewidth}|X|p{.058\linewidth}|p{.0595\linewidth}|p{.061\linewidth}|p{.061\linewidth}|X|X|}
    \hline
\cellcolor{jmotablecolor} Trial    & \cellcolor{jmotablecolor} \begin{tabular}[c]{@{}c@{}}FT-TBU\\ ID\end{tabular} & \cellcolor{jmotablecolor} \begin{tabular}[c]{@{}c@{}}$h_0'$  \\ ($\mu$m)\end{tabular} & \cellcolor{jmotablecolor} \cellcolor{jmotablecolor} \begin{tabular}[c]{@{}c@{}}$f_0'$  \\ (\%)\end{tabular} & \cellcolor{jmotablecolor} \begin{tabular}[c]{@{}c@{}}$v'$\\ ($\frac{\mu \text{m}}{\text{min}}$)\end{tabular} & \cellcolor{jmotablecolor} \begin{tabular}[c]{@{}c@{}}$a'$\\ ($s^{-1}$)\end{tabular} & \cellcolor{jmotablecolor} \begin{tabular}[c]{@{}c@{}}$b_1'$\\ ($s^{-1}$)\end{tabular} & \cellcolor{jmotablecolor} \begin{tabular}[c]{@{}c@{}}$b_2'$\\ ($s^{-1}$)\end{tabular} & \cellcolor{jmotablecolor} Norm & \cellcolor{jmotablecolor} Model \\ \hline
S9v1t4$^+$ & 4:00 --- & 3.32 & .324 & 24.1 & .0316 & .418 & 5.75 & .203 & M  \\ \hline
S9v2t1 & 3:00 --- & 5.01 & .292 & 27.3 & .461 & -.490 & .0715 & .110 & M  \\ \hline
S9v2t5 & 4:00 $\circ$ & 2.1 & .299 & 22.4 & .217 & -.417 & .882 & .118 & M \\ \hline
S9v2t5 & 4:30 $\circ$ & 2.33 & .299 & 50.9   & .360 & -.564 & .367 & .192   & M  \\ \hline
S10v1t6$^{++}$ & 12:30 $\circ$ & 3.08 & .293 & 1.27  &  & 1.95 & .277 & .0280 & D  \\ \hline
S13v2t10$^+$ & 6:30 --- & 3.59 & .259 & 26.4  &  & .138 & .102 & .121 & D  \\ \hline
S18v2t4$^{++}$ & 7:30 $\circ$ & 2.48 & .363 & 25.2  &  & 2.41 & 8.85 & .111 & D  \\ \hline
S27v2t2$^+$ & 5:00 --- & 1.91 & .4 & 9.32  & .714 & -.368  & .540 & .0271 & M  \\ \hline
\end{tabularx}
\caption{\footnotesize{Results from fitting various ODE models (up to four parameters). The subject (S) number, visit (v) number and (t) trial number are listed.  A $+$ denotes using a ``ghost'' first time level in the PDE fit and ``ghost'' time in the ODE fit.  The FT-TBU location is a clock reading taken from the center of the pupil.  FT-TBU type is denoted by --- for a streak, and $\circ$ for a spot. The initial TF thickness and FL concentration estimates are given. The optimal parameters are given for the case of the model with the smallest norm. The evaporative thinning rates are given by $v'$, constant extensional flow rate by $a'$ and decaying extensional flow and decay rates by $b_1'$ and $b_2'$.}}	
\label{table:results}
\end{table}

\subsection{Fitting examples}
\label{sec:examples}

\begin{figure}
\centering
\subfloat[][FL intensity with minima aligned]{
\includegraphics[scale=.15]{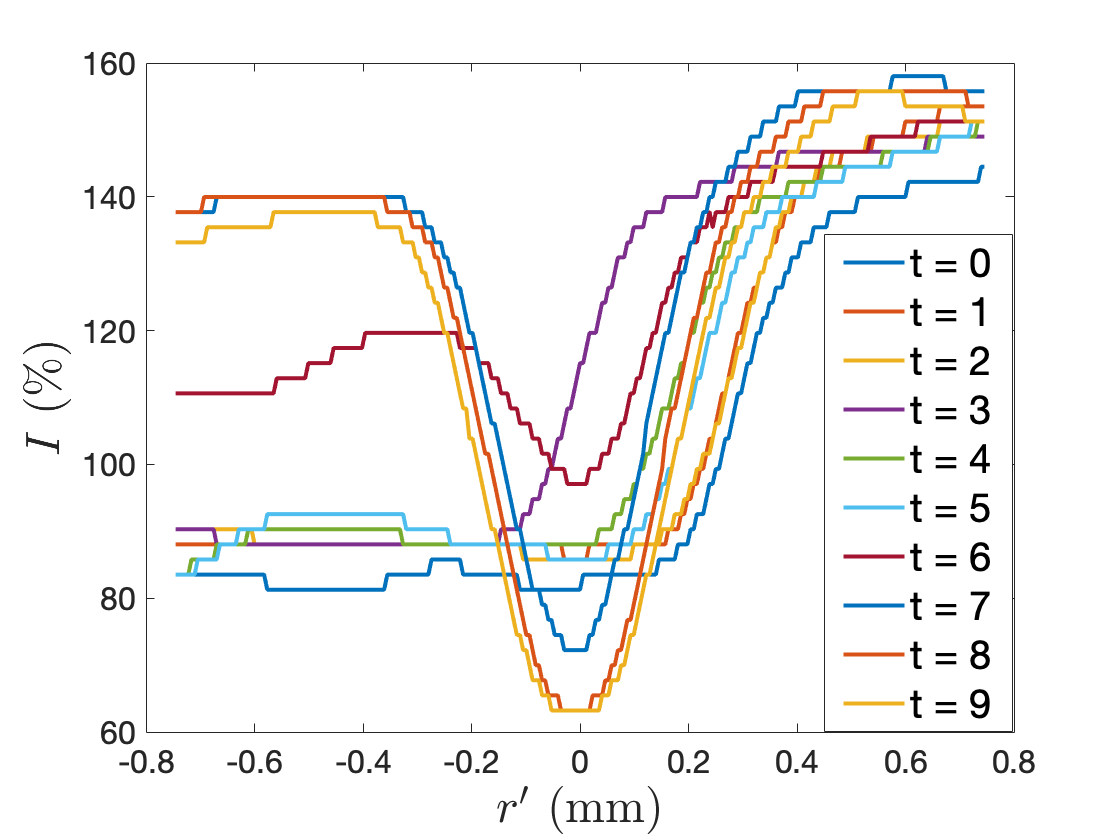}}
\subfloat[][FL intensity decrease]{
\includegraphics[scale=.15]{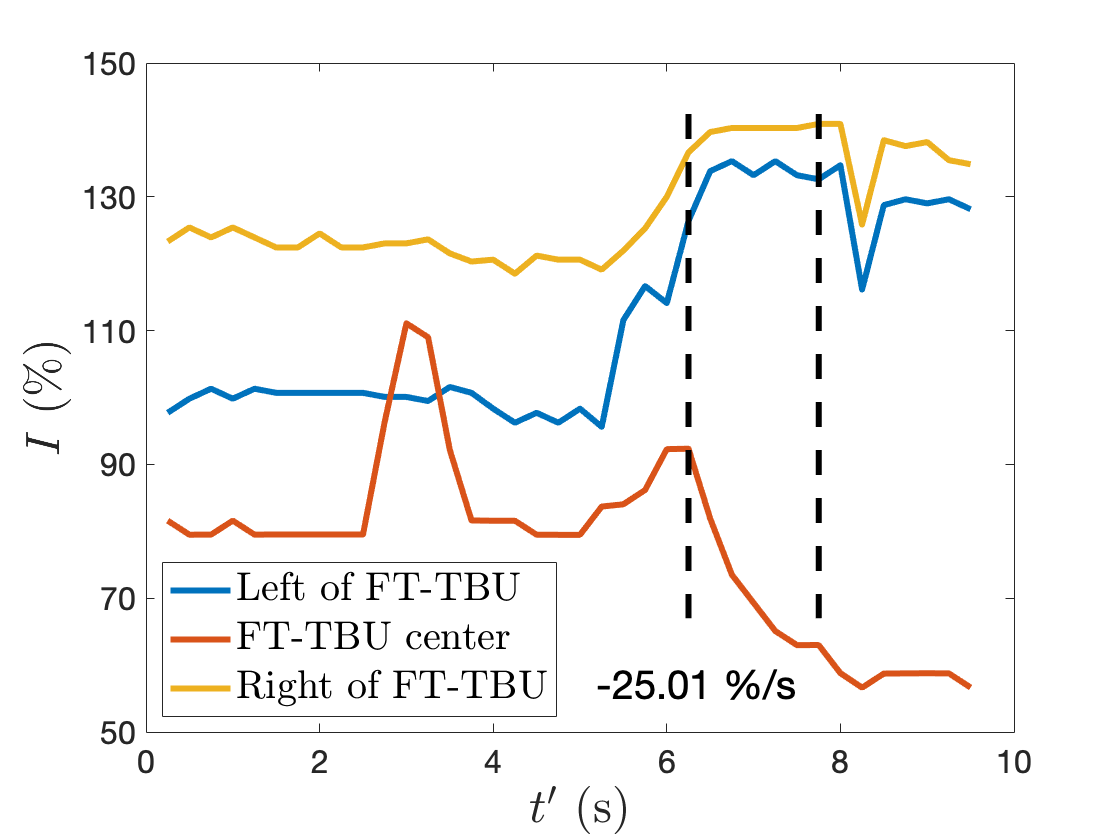}} \\
\hspace{8mm}
\subfloat[][FT-TBU data extraction]{
\includegraphics[scale=.12]{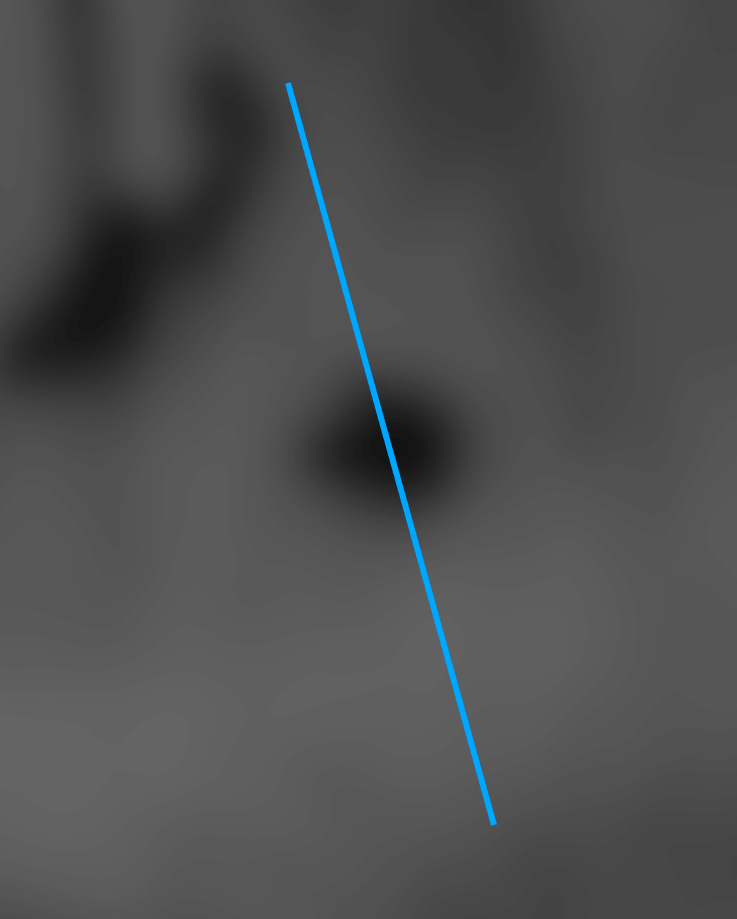}} \hspace{8mm}
\subfloat[][Exp. and best fit th. FL intensity]{\includegraphics[scale=.15]{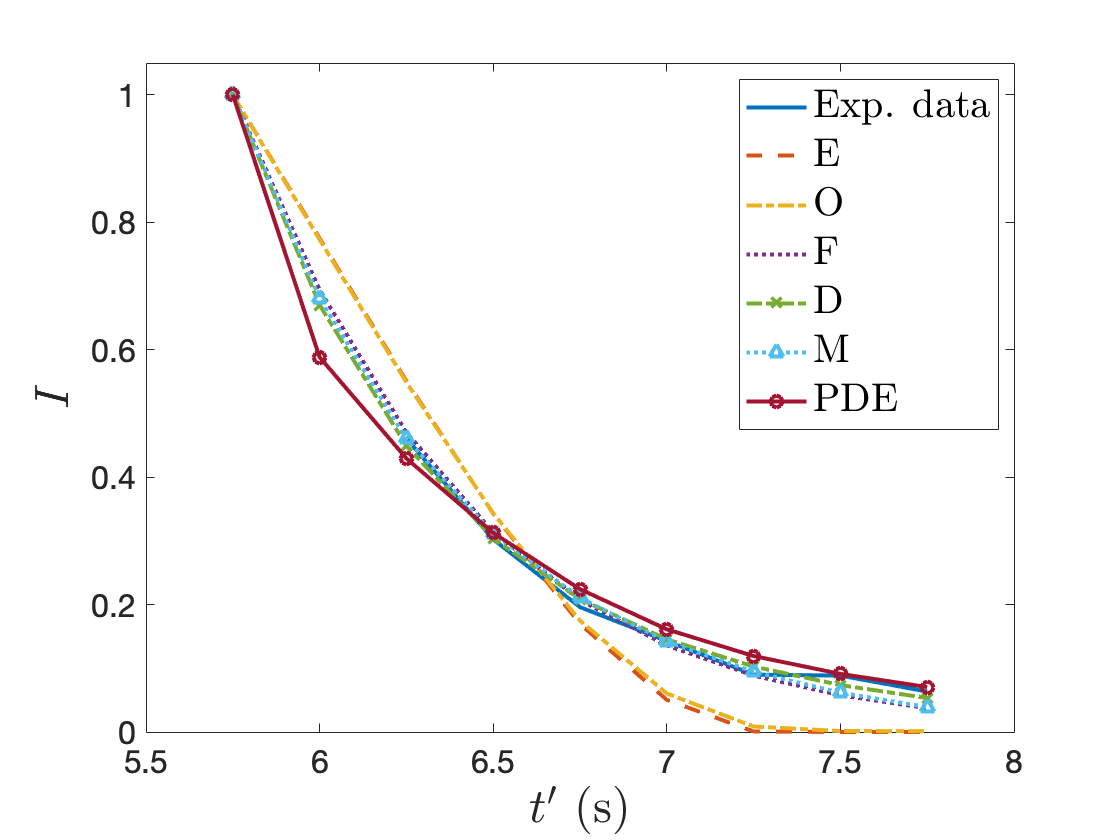}\label{fig:S10v1t6_fit}}
\caption{\footnotesize{Extracted data and best fit results for the S10v1t6 12:30 spot. In (c) the image has been brightened and contrast-enhanced. Case (c) evaporation (see Luke \textit{et al.}\cite{luke2021}) was used in the PDE fit.}}
\label{fig:S10v1t6_data}
\end{figure}

The S10v1t6 12:30 spot is shown as an example of our fitting procedure in Figure \ref{fig:S10v1t6_data}. Figure \ref{fig:S10v1t6_data}c shows the line of data extracted for the PDE fit recorded in Luke \textit{et al.}\cite{luke2021}; we fit the breakup data at the midpoint of the line with our ODE models. The results for each of the six ODE models are recorded in Table \ref{table:example}. In order to determine the model selected to report in Table \ref{table:results}, we compare the 2-norms of the difference between the theoretical and experimental FL intensities and select the case of the model corresponding to the smallest value.

The first six seconds of the trial are obscured by eyelashes and the upper eyelid. The spot has already started to form and darkens quickly after the breakup region is revealed around six seconds into the trial (see Figure \ref{fig:S10v1t6_data}b). In order to fit the data with our model, we use ``ghost'' time levels for 0.5 seconds. Figure \ref{fig:S10v1t6_fit} shows that the experimental FL intensity drops to less than 10\% of its initial value.

\begin{table}
\centering
\begin{tabularx}{\textwidth}{|p{.18\linewidth}|X|X|X|X|p{.15\linewidth}|p{.15\linewidth}|}
\hline
\cellcolor{jmotablecolor} Model & \cellcolor{jmotablecolor} \begin{tabular}[c]{@{}c@{}}$v'$\\ ($\frac{\mu \text{m}}{\text{min}}$)\end{tabular} & \cellcolor{jmotablecolor} \begin{tabular}[c]{@{}c@{}}$a'$\\ ($s^{-1}$)\end{tabular} & \cellcolor{jmotablecolor} \begin{tabular}[c]{@{}c@{}}$b_1'$\\ ($s^{-1}$)\end{tabular} & \cellcolor{jmotablecolor} \begin{tabular}[c]{@{}c@{}}$b_2'$\\ ($s^{-1}$)\end{tabular} & \cellcolor{jmotablecolor} Residual & \cellcolor{jmotablecolor} Norm \\ \hline
Evap only (E) & 120 &  &  & & 3.88 $\times 10^{-2}$ & 1.97 $\times 10^{-1}$ \\ \hline
Evap + osm (O) & 122 &  & & & 3.47 $\times 10^{-2}$ & 1.86 $\times 10^{-1}$ \\ \hline
Evap, osm, flow (F) & 0.00 & 1.74 & & & 2.0 $\times 10^{-3}$ & 4.50 $\times 10^{-2}$ \\ \hline
Evap, osm, dec. flow (D) & 1.27 & & 1.95 & 0.277  & 7.86 $\times 10^{-4}$ & 2.80 $\times 10^{-2}$ \\ \hline
Evap, osm, mixed flow (M) & 4.91 & 0.656  & 1.19 &  0.423 & 6.10 $\times 10^{-4}$ & 4.04 $\times 10^{-2}$ \\ \hline
Mixed-mech PDE center & 5.92 &  &  & &  & 5.58 $\times 10^{-2}$ \\ \hline
\end{tabularx}
\caption{\footnotesize{S10v1t6 12:30 center of spot data fit with the five cases of models. The central data of the best PDE fit is shown for comparison.}}
\label{table:example}
\end{table}

In Table \ref{table:example}, the two ODE models without flow select unrealistic evaporation rates in an attempt to match the rapid thinning of the S10v1t6 12:30 spot. On average, the evaporation rates chosen by the ODE models with flow are among the smallest optimal values for all mixed-mechanism fit instances. This is likely due to the relatively large flow rate parameters--unlike any other trial, the initial flow value $b_1'$ or $a'$ is above 1 $s^{-1}$ for each of the three models that involve flow. We take this as strong evidence that the Marangoni effect is the dominant mechanism causing the thinning. Further evidence of this statement is the fact that the case F model selected zero evaporation. This instance was fit well with a Marangoni effect-only PDE model which ignored evaporation, which is consistent with our small or zero optimal evaporation rates. The case D model produces the smallest residual. This FT-TBU instance exhibits the largest drop in FL intensity of all eight; the decaying model gives the best fit because the TF has likely thinned to almost zero thickness, allowing little flow, if any.

\begin{figure}
\centering
\subfloat[][FL intensity decrease]{
\includegraphics[scale=.15]{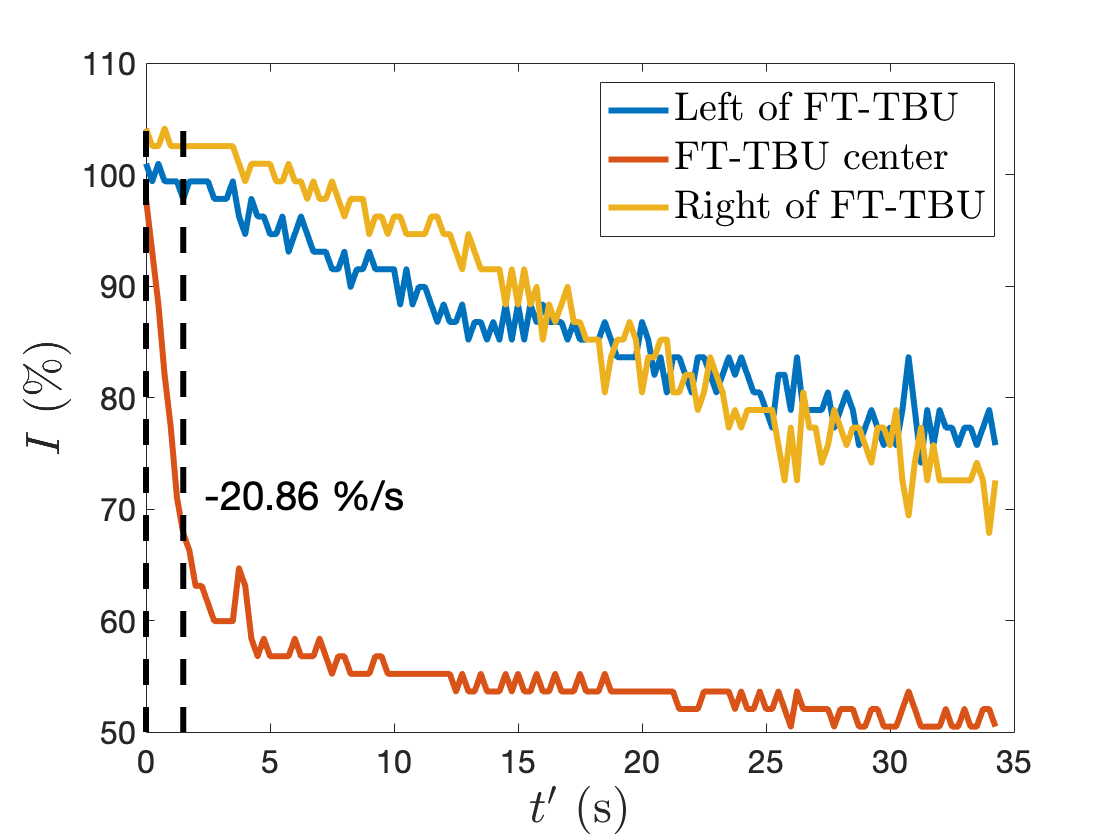}}
\subfloat[][Exp. and best fit th. FL intensity]{\includegraphics[scale=.15]{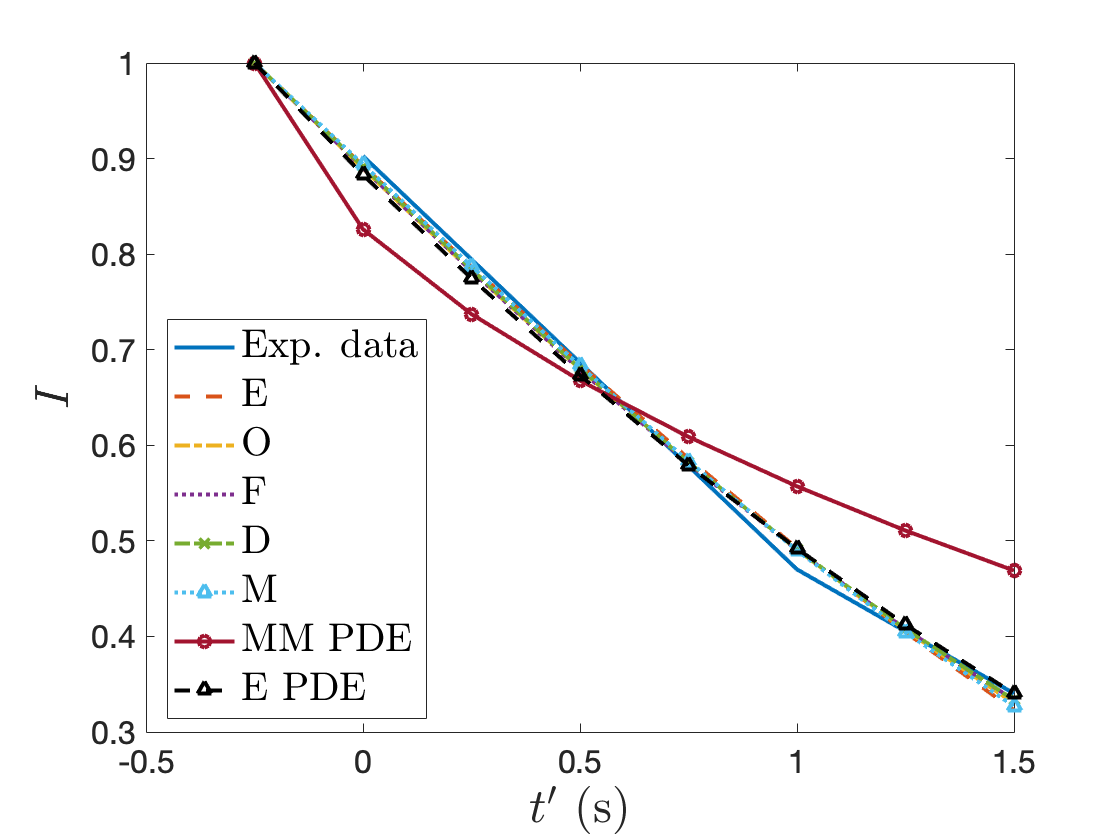}\label{fig:S27v2t2_5_fit}}
\caption{\footnotesize{Extracted data and best fit results for the S27v2t2 5:00 streak. Uniform evaporation was used in the mixed-mechanism PDE fit.}}
\label{fig:S27v2t2_data}
\end{figure}

The S27v2t2 5:00 streak data and fits are shown in Figure \ref{fig:S27v2t2_data}. As in Luke \textit{et al.}\cite{luke2021}, we use a quarter second of ``ghost'' time at the start of the fit. This instance is of particular interest because the center of the mixed-mechanism PDE theoretical FL intensity does not capture the dynamics of the experimental data well. In Luke \textit{et al.}\cite{luke2021}, the S27v2t2 5:00 streak was categorized as Marangoni-effect dominated due to the large Marangoni number and outward flow of the best fit.   The best fit case M model selects outward flow; however, the close second-best (D) and third-best (F) cases select inward flow. These latter two models also select a significantly larger evaporation rate than the others. The S27v2t2 5:00 streak was also fit with an evaporation-only model\cite{braun2018} in Luke \textit{et al.}\cite{luke2021}. That fit (E PDE) is shown along with the mixed-mechanism fit (MM PDE) and the ODE results in Figure \ref{fig:S27v2t2_5_fit} and outperforms the mixed-mechanism PDE fit. Further, the optimal peak evaporation rate for the evaporation-only PDE fit is 35.3 $\mu$m/min, which is a large but plausible evaporation rate. This suggests that evaporation may play a larger role in this instance than previously thought.

\begin{figure}
\centering
\subfloat[][FL intensity decrease]{
\includegraphics[scale=.15]{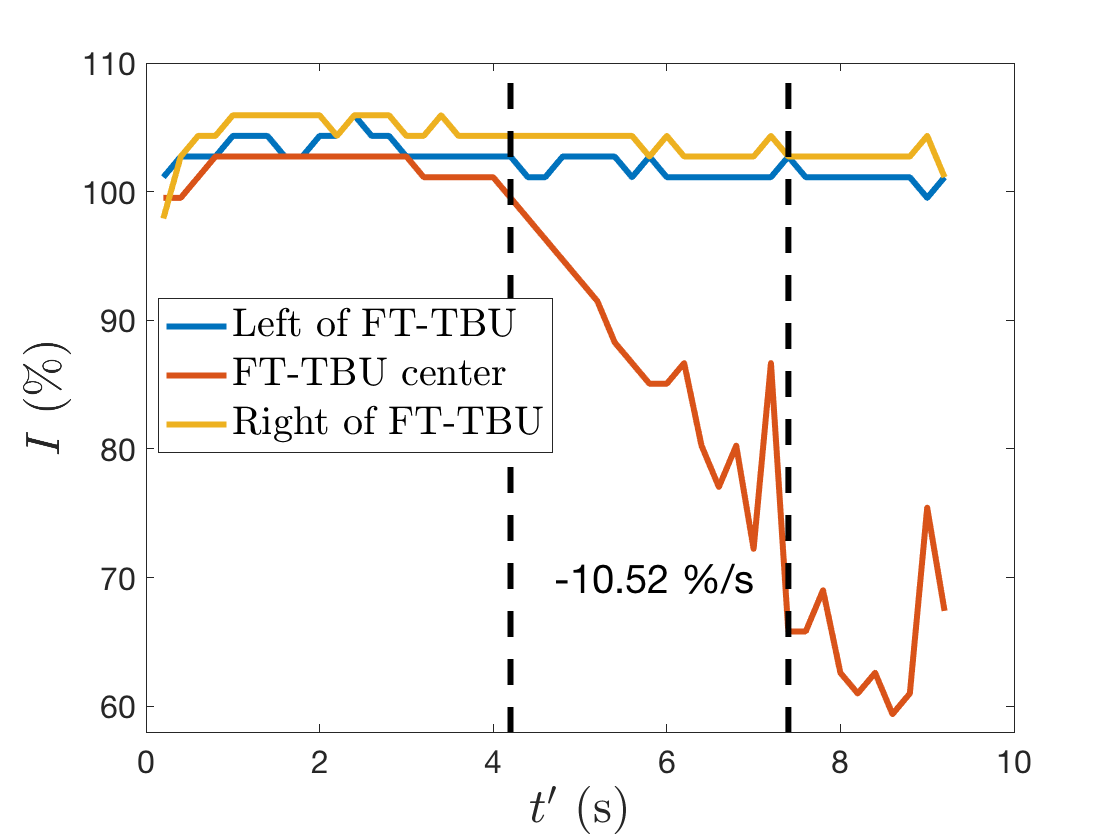}}
\subfloat[][Exp. and best fits th. FL intensity]{\includegraphics[scale=.15]{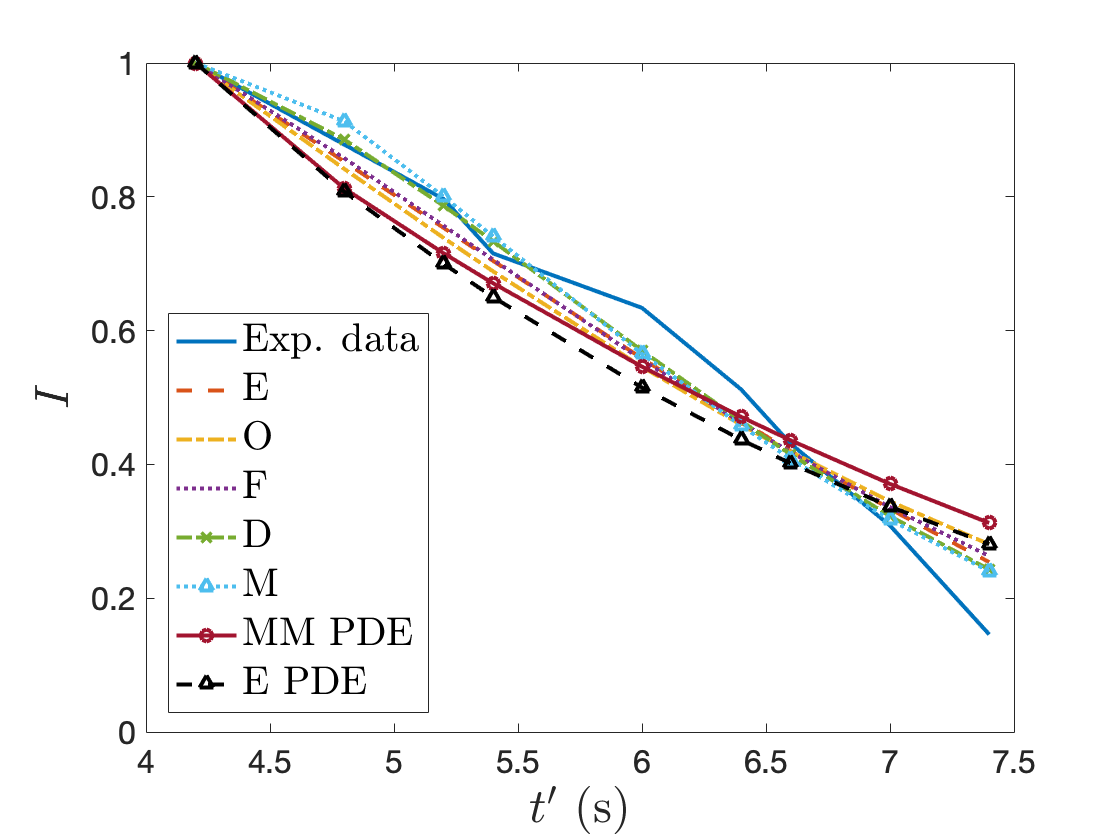}
\label{fig:S9v2t5_fit}}
\caption{\footnotesize{Extracted data and best fit results for the S9v2t5 4:00 spot data. Case (c) evaporation (see Luke \textit{et al.}\cite{luke2021}) was used in the mixed-mechanism PDE fit.}}
\label{fig:S9v2t5_data}
\end{figure}

In Figures \ref{fig:S9v2t5_data}a-c we show the S9v2t5 4:00 spot data and in Figure \ref{fig:S9v2t5_fit} we show the fits. We have plotted the central data from both the best-fit mixed-mechanism (MM PDE) and evaporation-only (E PDE) models for comparison because this instance is also fit well with the latter model and was categorized as evaporation-dominated in Luke \textit{et al.}\cite{luke2021}. All three ODE models with flow select some amount of inward flow, which aligns well with the PDE model, whose flow profile changes sign at the origin as time progresses (see Table \ref{tab:dubardr}). In all cases, the flow is of a significantly smaller magnitude than the Marangoni effect-dominated or transitional thinning instances. The case M model gives the smallest residual. Notably, the evaporation-only fits for this instance give closer residuals to the best fit model than other instances; this suggests that evaporation is the dominant mechanism causing the thinning.

\section{Discussion}
\label{sec:disc}

The quantities recorded in Table \ref{table:results} show more variation than the PDE results in some cases, but the qualitative similarities in the solutions are an important takeaway. For each TBU instance, the best fit ODE model includes time-dependent flow. This is strong evidence that evaporation alone cannot explain this thinning and that the Marangoni effect played a role, since it is characterized by non-constant thinning.  

\begin{figure}
\centering
\subfloat[][S10v1t6 12:30 spot]{\includegraphics[scale=.15]{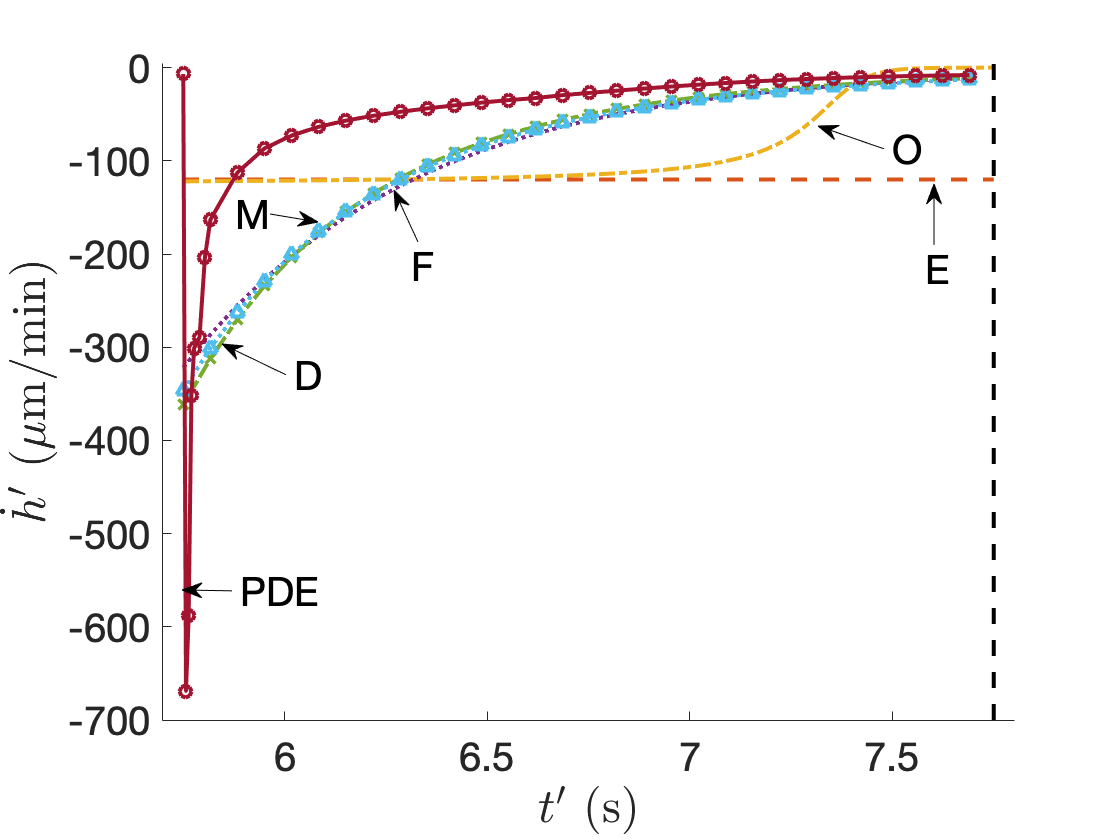}}
\subfloat[][S27v2t2 5:00 streak]{\includegraphics[scale=.149]{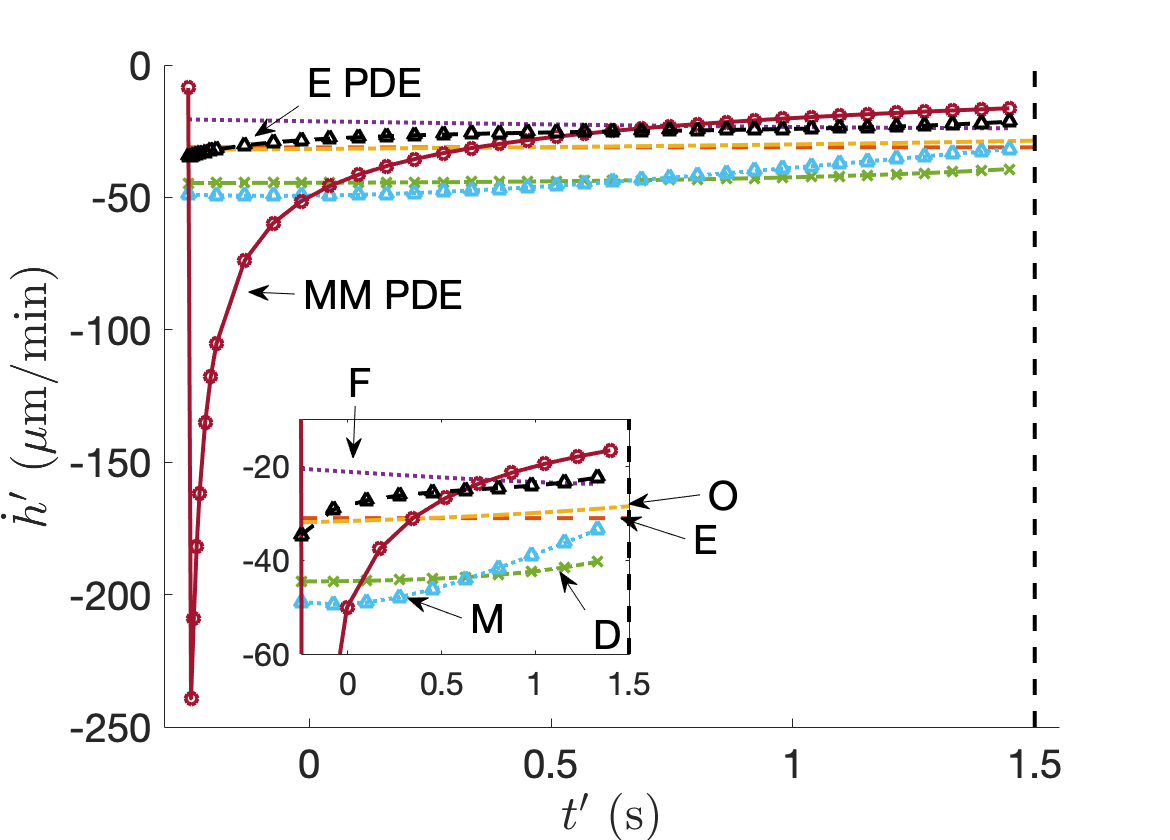}} \\
\subfloat[][S9v2t5 4:00 spot]{\includegraphics[scale=.15]{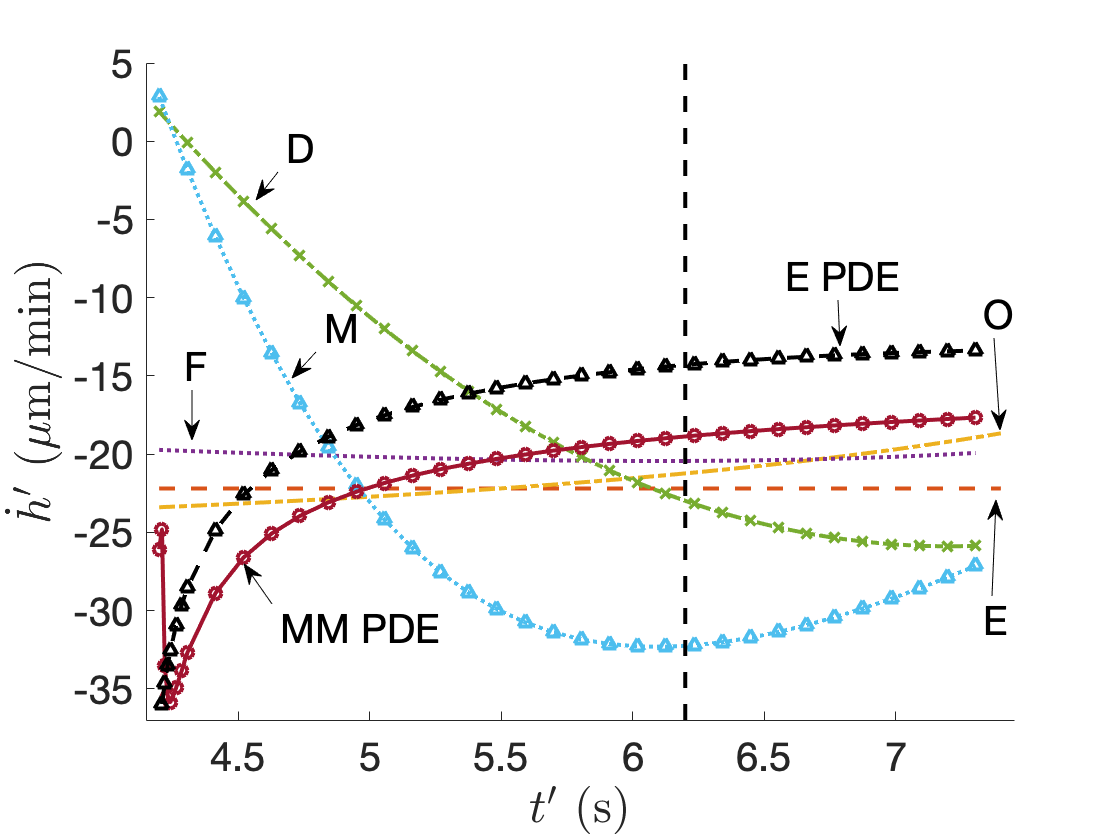}}
\caption{\footnotesize{$\dot{h}'$ from the five ODE models are plotted alongside $\partial_{t'} h'$ from the PDE model. The time point at which averaging will begin is shown as a dashed vertical line.}}
\label{fig:dhdt}
\end{figure}

In Figure \ref{fig:dhdt} we show the various time derivatives $\dot{h}'$ computed from the optimal values of the ODE models as well as the optimal $\partial_{t'}h'$ measured at the origin for the three examples of mixed-mechanism fitting shown in Section \ref{sec:examples}. The average starting two seconds in or the value at the final time point is recorded in Table \ref{tab:dhdt}. This delay in averaging matches the approach in Luke \textit{et al.}\cite{luke2021} and mimics experimental procedures\cite{nichols2005}. These values are shown along with the optimal evaporation rates for comparison. The S10v1t6 12:30 spot, which showed the most rapid thinning when fit with the mixed-mechanism PDE model, shows dynamic rates of thinning for a large portion of the trial for many of the ODE fits in Figure \ref{fig:dhdt}a. The case (D) and (M) ODE model $\dot{h}'$ values are very close, which we expect since they gave similar residuals when fit to the data. The S9v2t5 4:00 spot shows non-constant dynamics near the end of the trial in Figure \ref{fig:dhdt}c and we see further qualitative and quantitative agreement between the mixed-mechanism and evaporation-only PDE results. The case (D) and (M) models for this instance also show $\dot{h}' > 0$ in the first quarter second; the theoretical TF thickness solution is in fact slightly positive early on. This is likely an attempt by the optimization to fit the concave down portion of the data in the first second or so. In general, the PDE models produce $\partial_{t'}h'$ values in the first quarter second that are much larger than the corresponding ODE numbers.

In Table \ref{tab:dhdt} we record the optimal evaporation rate (the top number) and an average thinning rate (the bottom number) for the PDE fit and each of the five ODE fits. The average thinning rate is either taken starting two seconds into the trial, or if the trial is two seconds or less, the value at the final time is recorded. The values corresponding to the best fit by an ODE model as determined by the smallest residual are shaded. In all instances, the overall thinning rate of the best ODE fit is larger than that of the PDE. This may reflect the tendency of the theoretical PDE FL intensity to lag the experimental FL intensity in later times. In most instances, the best fit overall thinning rate is larger than the evaporation rate, indicative of outward thinning that supports the notion that the Marangoni effect contributed to the thinning. Notably, for the S9v2t5 4:30 spot, which was categorized as evaporation-dominated in Luke \textit{et al.}\cite{luke2021}, the evaporation rate is larger than the thinning rate, suggesting inward capillary flow combats the thinning. Some short trials exhibit rapid dynamics which occur early on; the recorded thinning rate may not represent the entirety of the trial.

 \begin{table}
\centering
\begin{tabularx}{\linewidth}{|X|X|X|X|X|X|X|X|X|X|}
\hline
\cellcolor{jmotablecolor} Trial & \cellcolor{jmotablecolor} \begin{tabular}[c]{@{}c@{}}FT-TBU\\ ID\end{tabular}  & \cellcolor{jmotablecolor} \begin{tabular}[c]{@{}c@{}} $v'$ \\ $\partial_{t'} h' $  \end{tabular} &  \cellcolor{jmotablecolor} \begin{tabular}[c]{@{}c@{}} $v'_E$ \\ $\dot{h'}$  \end{tabular} & \cellcolor{jmotablecolor} \begin{tabular}[c]{@{}c@{}} $v'_{O}$ \\ $\dot{h'}$ \end{tabular} & \cellcolor{jmotablecolor} \begin{tabular}[c]{@{}c@{}} $v'_F$ \\ $\dot{h'}$  \end{tabular} & \cellcolor{jmotablecolor} \begin{tabular}[c]{@{}c@{}} $v'_{D}$ \\ $\dot{h'}$  \end{tabular} & \cellcolor{jmotablecolor} \begin{tabular}[c]{@{}c@{}} $v'_{M}$ \\ $\dot{h'}$  \end{tabular} \\ \hline

S9v1t4 & 4:00 ---  & \begin{tabular}[c] {@{}c@{}} -6.26 \\ -15.8 \end{tabular} & \begin{tabular}[c] {@{}c@{}} -28.7 \\ -28.7 \end{tabular} & \begin{tabular}[c] {@{}c@{}} -30.2 \\ -26.4 \end{tabular} & \begin{tabular}[c] {@{}c@{}} -13.5 \\ -26.3 \end{tabular}& \begin{tabular}[c] {@{}c@{}} -16.4  \\ -26.0 \end{tabular}&\cellcolor[HTML]{C0C0C0} \begin{tabular}[c] {@{}c@{}} -24.1 \\ -24.0 \end{tabular}  \\ \hline

S9v2t1 & 3:00 ---  & \begin{tabular}[c] {@{}c@{}} -30.3 \\ -21.2 \end{tabular}
& \begin{tabular}[c] {@{}c@{}}-30.4 \\ -30.4 \end{tabular}
& \begin{tabular}[c] {@{}c@{}} -31.9 \\-28.5 \end{tabular}
& \begin{tabular}[c] {@{}c@{}} -29.1 \\ -29.2 \end{tabular}
& \begin{tabular}[c] {@{}c@{}} -38.6 \\ -20.6 \end{tabular}
& \cellcolor[HTML]{C0C0C0} \begin{tabular}[c] {@{}c@{}} -27.3 \\ -38.1 \end{tabular} \\ \hline

S9v2t5 & 4:00 $\circ$  & \begin{tabular}[c] {@{}c@{}} -26.2 \\ -18.2 \end{tabular}& \begin{tabular}[c] {@{}c@{}} -22.2 \\ -22.2 \end{tabular}&  \begin{tabular}[c] {@{}c@{}}-23.4 \\ -20.1 \end{tabular}& \begin{tabular}[c] {@{}c@{}} -27.7 \\ -20.2 \end{tabular}& \begin{tabular}[c] {@{}c@{}} -37.5  \\ -25.1 \end{tabular}& \cellcolor[HTML]{C0C0C0}\begin{tabular}[c] {@{}c@{}} -22.4 \\ -30.0 \end{tabular}\\ \hline

S9v2t5 & 4:30 $\circ$  & \begin{tabular}[c] {@{}c@{}} -36.9 \\ -23.4 \end{tabular}& \begin{tabular}[c] {@{}c@{}} -42.8 \\ -42.8 \end{tabular}& \begin{tabular}[c] {@{}c@{}} -44.6 \\ -35.7 \end{tabular}& \begin{tabular}[c] {@{}c@{}} -50.0 \\ -36.5 \end{tabular}& \begin{tabular}[c] {@{}c@{}} -48.7 \\ -38.2 \end{tabular}& \cellcolor[HTML]{C0C0C0} \begin{tabular}[c] {@{}c@{}} -50.9 \\ -44.9  \end{tabular}\\ \hline

S10v1t6 & 12:30 $\circ$  & \begin{tabular}[c] {@{}c@{}} -5.92 \\ -7.60\end{tabular} & \begin{tabular}[c] {@{}c@{}} -120 \\ -120 \end{tabular} & \begin{tabular}[c] {@{}c@{}} -122 \\ -0.0131 \end{tabular}& \begin{tabular}[c] {@{}c@{}} 0 \\ -9.92\end{tabular} &\cellcolor[HTML]{C0C0C0} \begin{tabular}[c] {@{}c@{}} -1.27 \\ -10.4 \end{tabular} & \begin{tabular}[c] {@{}c@{}} -4.91 \\ -11.5 \end{tabular} \\ \hline

S13v2t10 & 6:30 ---   & \begin{tabular}[c] {@{}c@{}} -13.6 \\ -21.8 \end{tabular} & \begin{tabular}[c] {@{}c@{}} -37.1 \\ -37.1 \end{tabular} & \begin{tabular}[c] {@{}c@{}} -38.9  \\ -33.6 \end{tabular} & \begin{tabular}[c] {@{}c@{}} -21.5 \\ -31.7 \end{tabular} &\cellcolor[HTML]{C0C0C0} \begin{tabular}[c] {@{}c@{}} -26.4 \\ -31.1 \end{tabular} & \begin{tabular}[c] {@{}c@{}} -20.4 \\ -30.8 \end{tabular} \\ \hline

S18v2t4 & 7:30 $\circ$  & \begin{tabular}[c] {@{}c@{}} -13.1 \\ -16.3 \end{tabular}& \begin{tabular}[c] {@{}c@{}} -37.1 \\  -37.1 \end{tabular}& \begin{tabular}[c] {@{}c@{}} -39.0 \\  -32.4 \end{tabular} & \begin{tabular}[c] {@{}c@{}} -0.0011 \\ -18.9 \end{tabular} &\cellcolor[HTML]{C0C0C0} \begin{tabular}[c] {@{}c@{}} -25.2 \\ 21.0 \end{tabular}& \begin{tabular}[c] {@{}c@{}} -20.3 \\ -20.5 \end{tabular} \\ \hline

S27v2t2 & 5:00 --- & \begin{tabular}[c] {@{}c@{}} -6.11 \\ -16.0 \end{tabular} & \begin{tabular}[c] {@{}c@{}} -31.1 \\ -31.1 \end{tabular}& \begin{tabular}[c] {@{}c@{}} -32.0 \\ -28.6 \end{tabular}& \begin{tabular}[c] {@{}c@{}} -43.6 \\ -23.9 \end{tabular}& \begin{tabular}[c] {@{}c@{}} -47.8 \\ -38.8 \end{tabular}&\cellcolor[HTML]{C0C0C0} \begin{tabular}[c] {@{}c@{}} -9.32 \\ -30.9 \end{tabular}\\ \hline
\end{tabularx}
\caption{\footnotesize{The optimal evaporation rates are recorded along with estimates of average $\dot{h'}$ for the mixed-mechanism model fits (starting two seconds into the trial). All rates are measured in $\mu$m/min. PDE values are given by $v'$ and $\partial_{t'}h'$; the rest are from the various ODE models. The value at the last time was used for trials less than two seconds in length. The five cases of ODE models are in order and denoted by subscripts on the evaporation value: $v'_E$, $v_{O}'$, $v_F'$, $v_D'$ and $v_{M}'$. The shaded entries correspond to the model giving the best fit as determined by the smallest norm.}}
\label{tab:dhdt}
\end{table}

\begin{figure}
\centering
\subfloat[][S10v1t6 12:30 spot]{\includegraphics[scale=.15]{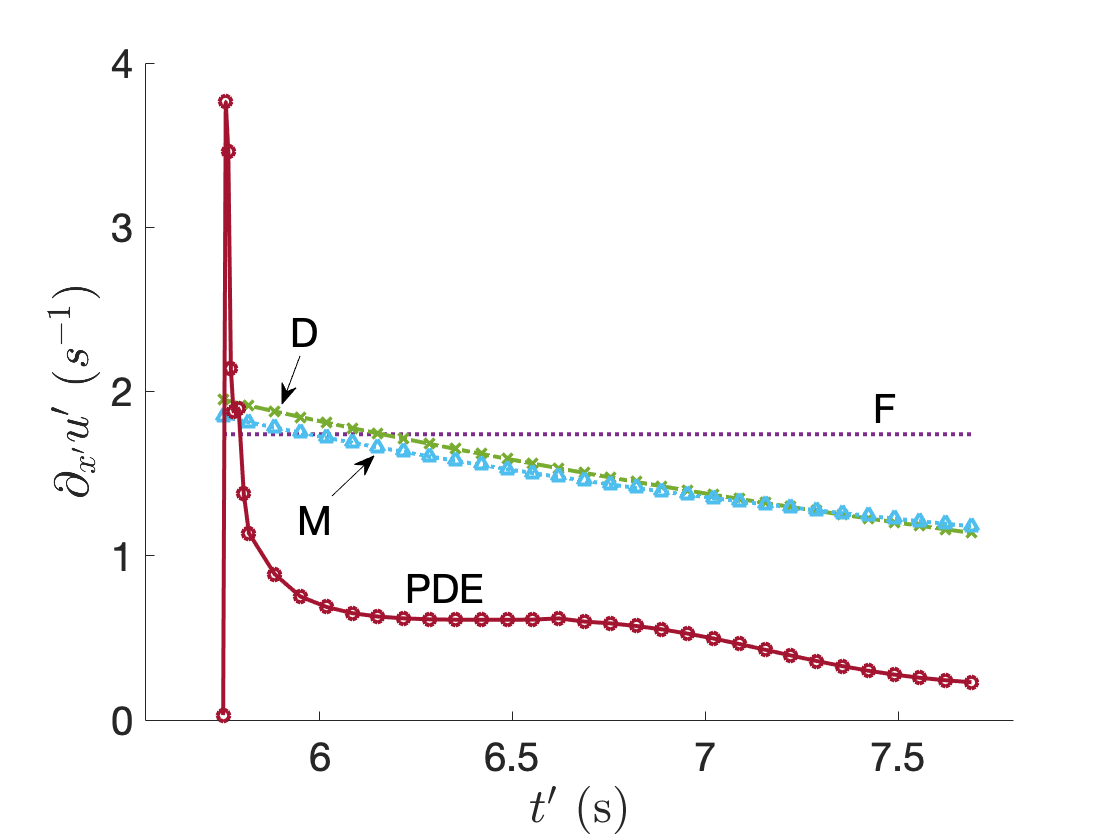}}
\subfloat[][S27v2t2 5:00 streak]{\includegraphics[scale=.15]{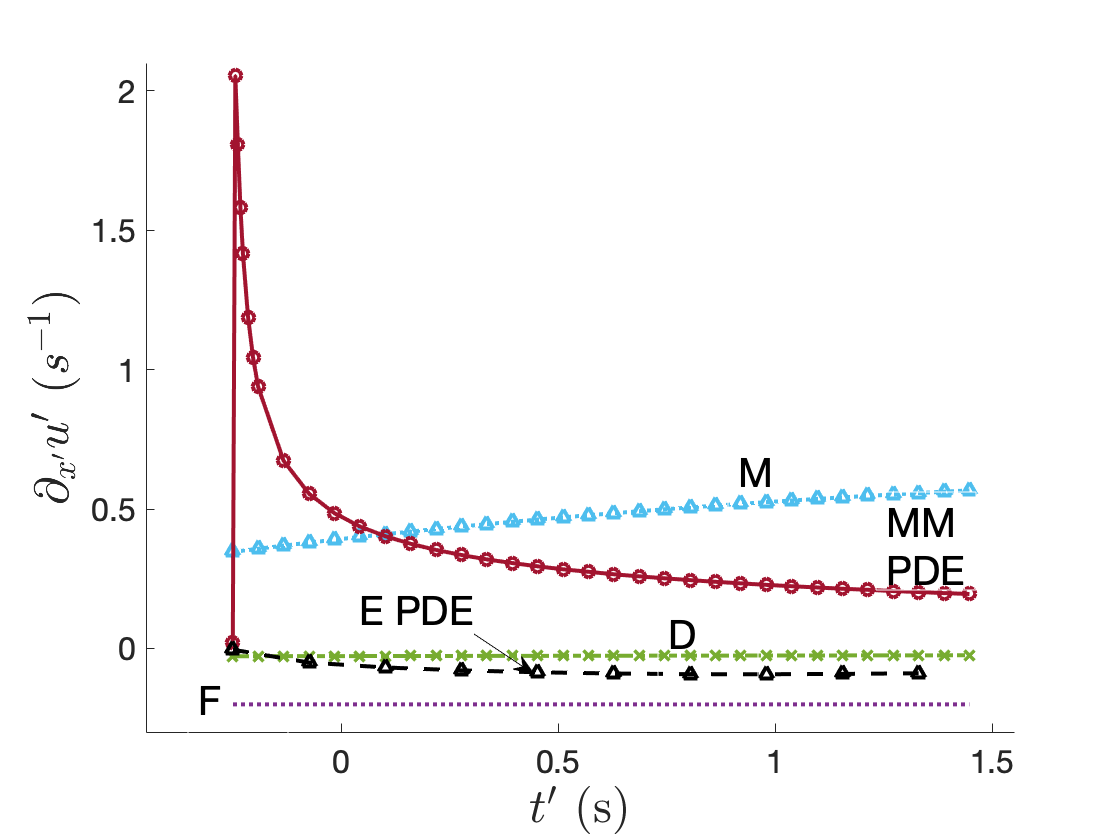}} \\
\subfloat[][S9v2t5 4:00 spot]{\includegraphics[scale=.15]{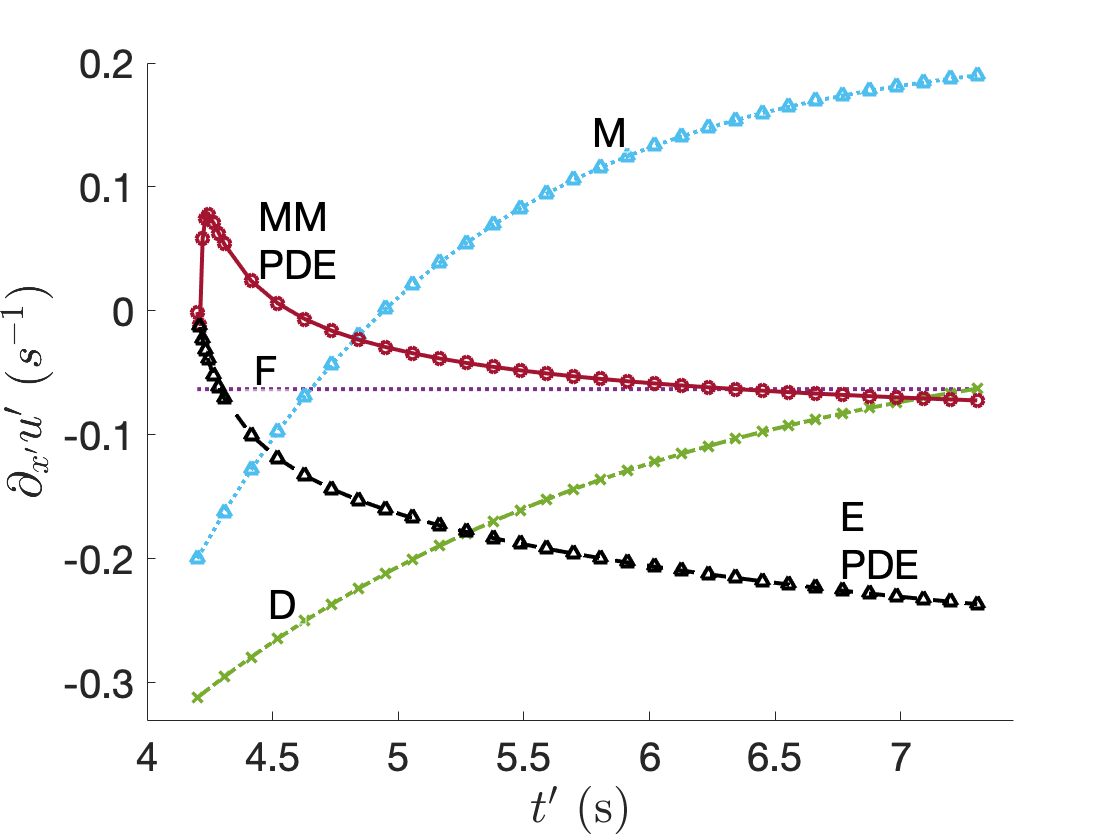}}
\caption{\footnotesize{The mixed-mechanism $\partial_{r'} \bar{u}'$ or $\partial_{x'} \bar{u}'$ are plotted against the strain rate $\partial_{x'}u'$ for each ODE model with flow.}}
\label{fig:dubardr}
\end{figure}

We plot $\partial_{r'} \bar{u}'$ or $\partial_{x'} \bar{u}'$ for the three examples from Section \ref{sec:examples} in Figure \ref{fig:dubardr} along with $\partial_{x'} u'$ from the relevant ODE models. In Figures \ref{fig:dubardr}b,c we also plot the evaporation-only PDE profiles. At least one ODE model does a decent job approximating the qualitative behavior of the PDE flow profile after the first quarter second or so. A notable exception is the constant extensional flow option for the S10v1t6 12:30 spot; this suggests the flow profile is highly time-dependent. An average value for each model and instance is taken over the whole trial and recorded in Table \ref{tab:dubardr}.

In Table \ref{tab:dubardr} we record the flow profile data for each instance. The shaded entries correspond to the best fit. From left to right, the parameters correspond to the coefficients of the flow terms from the case F, D, and M models, respectively. For most cases, the optimal flow directions match the PDE results, an indicator that they were correctly classified in Luke \textit{et al.}\cite{luke2021}. The three instances classified as evaporation-dominated or transitional thinning in Luke \textit{et al.}\cite{luke2021} show some amount of inward flow in both the PDE and ODE best fits, which is consistent since evaporation-dominated thinning is characterized by inward flow. In the cases of the S9v1t4 4:00 streak, S10v1t6 12:30 spot, S13v2t10 6:30 streak, and S18v2t4 7:30 spot, the ODE strain rates are always positive, which match the signs of the average strain rates of their corresponding PDE fits. This is evidence that outward tangential flow is important for explaining these instances, and so the thinning is likely influenced by the Marangoni effect. In the case of the S27v2t2 5:00 streak, the final strain rate signs of the best fit ODE and mixed-mechanism PDE models do not match. In this instance, the evaporation-only and mixed-mechanism PDEs show qualitative differences, and the evaporation-only PDE gives a better fit to the central data than the mixed-mechanism version. Perhaps the breakup dynamics of the streak in fact includes inward flow. For the S27v2t2 5:00 streak, this theory is supported by the text in Section \ref{sec:examples}.

\begin{table}
\centering
\begin{tabularx}{\linewidth}{|p{.09\linewidth}|p{.08\linewidth}|p{.08\linewidth}|p{.08\linewidth}|p{.08\linewidth}|X|p{.08\linewidth}|X|X|}
\hline
\cellcolor{jmotablecolor} Trial & \cellcolor{jmotablecolor} \begin{tabular}[c]{@{}c@{}}FT-TBU\\ ID\end{tabular} &  \cellcolor{jmotablecolor} \begin{tabular}[c]{@{}c@{}}$\dot{\gamma}'$ \end{tabular} & \cellcolor{jmotablecolor} \begin{tabular}[c]{@{}c@{}} $a'_F$  \end{tabular} & \cellcolor{jmotablecolor} \begin{tabular}[c]{@{}c@{}}  $b_{1D}'$ \end{tabular} &  \cellcolor{jmotablecolor} \begin{tabular}[c]{@{}c@{}}  $b_{2D}'$ \end{tabular} & \cellcolor{jmotablecolor} \begin{tabular}[c]{@{}c@{}}  $a_M'$  \end{tabular} &  \cellcolor{jmotablecolor} \begin{tabular}[c]{@{}c@{}}  $b_{1M}'$ \end{tabular}  & \cellcolor{jmotablecolor} \begin{tabular}[c]{@{}c@{}} $b_{2M}'$  \end{tabular}   \\ \hline
S9v1t4 & 4:00 --- & .150 & .189 & .164 & .0421 &\cellcolor[HTML]{C0C0C0} .0316 &\cellcolor[HTML]{C0C0C0} .418 & \cellcolor[HTML]{C0C0C0} 5.75 \\ \hline

S9v2t1 & 3:00 --- & -.0218 & .0199 & -.0894 & .385  &\cellcolor[HTML]{C0C0C0} .461 &\cellcolor[HTML]{C0C0C0} -.490  &\cellcolor[HTML]{C0C0C0} .0715 \\ \hline

S9v2t5 & 4:00 $\circ$ & -.0427   & -.0631 & -.312 & .517 & \cellcolor[HTML]{C0C0C0} .217 &\cellcolor[HTML]{C0C0C0} -.417 & \cellcolor[HTML]{C0C0C0} .882 \\ \hline

S9v2t5 & 4:30 $\circ$ & -.0786 & -.0757 & -.963 & .708 &\cellcolor[HTML]{C0C0C0} .360 &\cellcolor[HTML]{C0C0C0} -.564 & \cellcolor[HTML]{C0C0C0} .367  \\ \hline

S10v1t6 & 12:30 $\circ$ & .572  & 1.74 &\cellcolor[HTML]{C0C0C0} 1.95 &\cellcolor[HTML]{C0C0C0} .277 & .656 & 1.19 & .423  \\ \hline

S13v2t10 & 6:30 --- & .147 & .172 &\cellcolor[HTML]{C0C0C0} .138 &\cellcolor[HTML]{C0C0C0} .102 & .173 & .0257 & .812 \\ \hline

S18v2t4 & 7:30 $\circ$ & .172  & .674 & \cellcolor[HTML]{C0C0C0} 2.41 &\cellcolor[HTML]{C0C0C0} 8.85 & .0733 & 3.69 & 12.8 \\ \hline

S27v2t2 & 5:00 --- & .343  & -.201 & -.282 &  .0761 & \cellcolor[HTML]{C0C0C0} .714 &\cellcolor[HTML]{C0C0C0} -.368 & \cellcolor[HTML]{C0C0C0} .540 \\ \hline

\end{tabularx}
\caption{\footnotesize{Estimates of the extensional rate $\dot{\gamma}'$, which is either $\partial_{r'} \bar{u}'$ for spots or $ \partial_{x'} \bar{u}'$ for streaks, at the origin taken over the entire trial length in ($s^{-1}$) for the mixed-mechanism model fits. These are compared with the optimal values from the three ODE models with flow. The shaded entries correspond to the model giving the best fit as determined by the smallest norm.}}
\label{tab:dubardr}
\end{table}

In general, the ODE models do a good job of capturing the essence of the dynamics of the PDE model fits. Each instance that we expect to have some outward flow has at least one positive flow parameter, and vice versa for the inward flow instances (see Table \ref{tab:dubardr}). The ODE models are able to be a little more fine-tuned, as the PDE fit is done over space as well; as such, we might expect the central PDE data that is shown for comparison not to match the data quite as closely. The PDE fits often struggle to keep up with the experimental data in later times; this is reflected in the slower (in general) average thinning rate $\partial_{t'} h'$ as compared to the ODE values $\dot{h}'$ (see Table \ref{tab:dhdt}). However, the ODE data and fit is a simplification of the overall breakup dynamics, and viewing temporospatial data has value on its own.

\begin{figure}
\centering
\subfloat[][Evaporation rates]{\includegraphics[scale=.16]{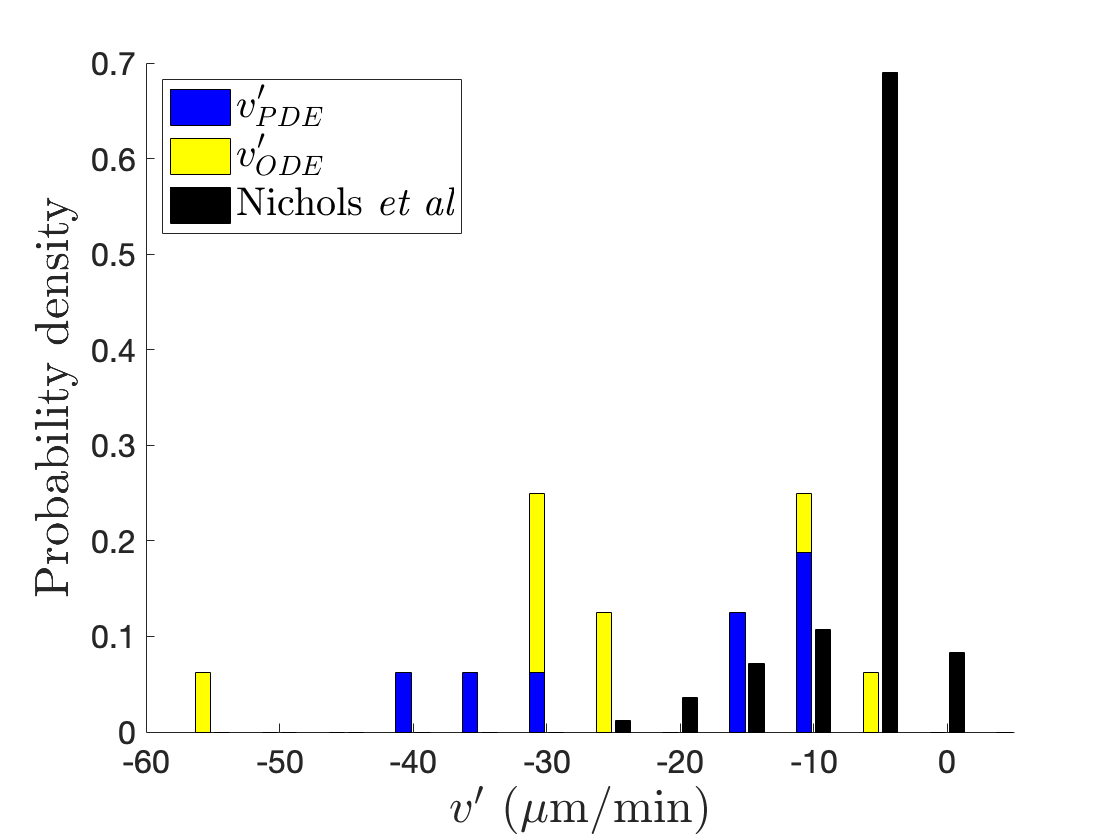}\label{fig:thin_hist_e}}
\subfloat[][Overall thinning rates]{\includegraphics[scale=.16]{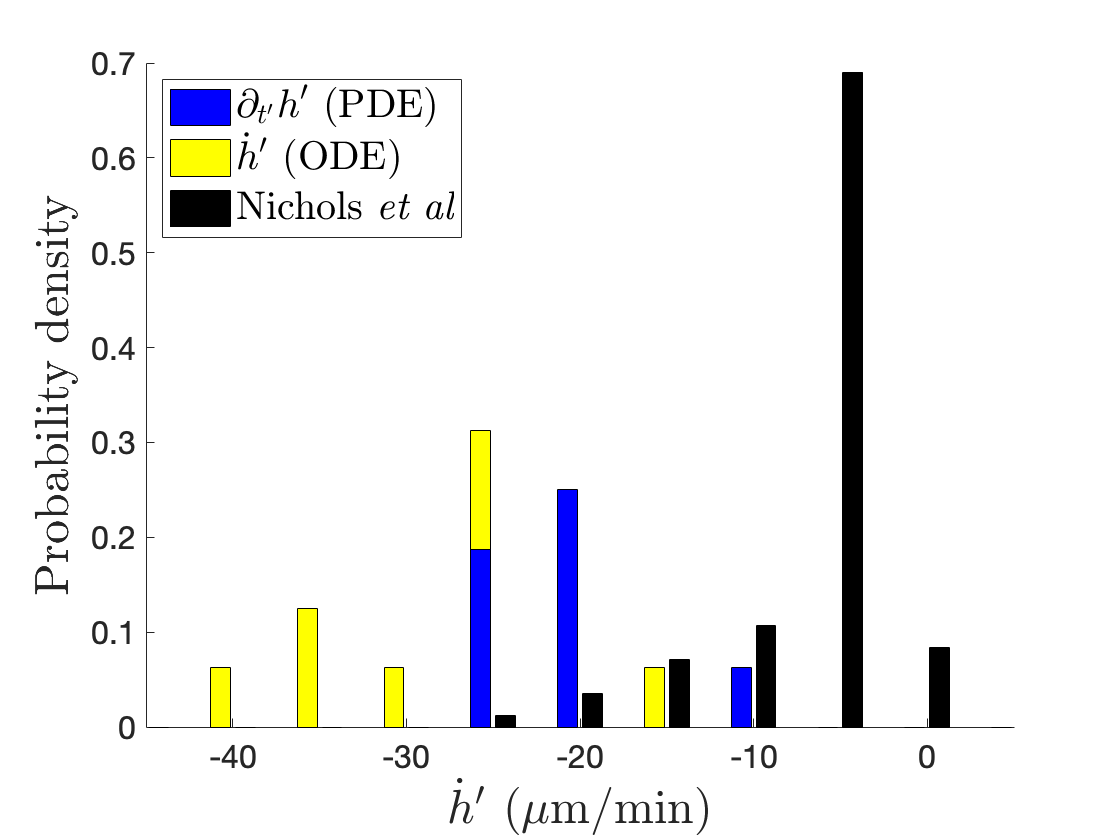}\label{fig:thin_hist_o}}
\caption{\footnotesize{Histograms of rates of change plotted against experimental point measurements from Nichols \textit{et al.}\cite{nichols2005}; note that the experiment cannot distiguish between $v'$ and $\dot{h}'$. The best fit ODE model data is shown as determined by the smallest norm.}}
\label{fig:thin_hist}
\end{figure}

Figure \ref{fig:thin_hist} compares the PDE and ODE best fit evaporation and thinning rate results to experimental point measurements reported in Nichols \textit{et al.}\cite{nichols2005}. For both histograms, a bin size of 5 $\mu$m/min was used. While the ODE rates show a wider range than the PDE results, there is significant overlap. The overall thinning rate is more comparable to the Nichols \textit{et al.}\cite{nichols2005} data since that study could not separate evaporation from the other mechanisms affecting TF thickness. We expect our thinning rates to be larger than the point measurements since the Nichols \textit{et al.}\cite{nichols2005} study did not target breakup. While some of the ODE data lies outside the experimental distribution, many ODE thinning rate values are comparable, suggesting these simplified models return physically relevant quantities that cannot be otherwise estimated. 

\begin{figure}
\centering
\subfloat[][Maximum osmolarity]{\includegraphics[scale=.16]{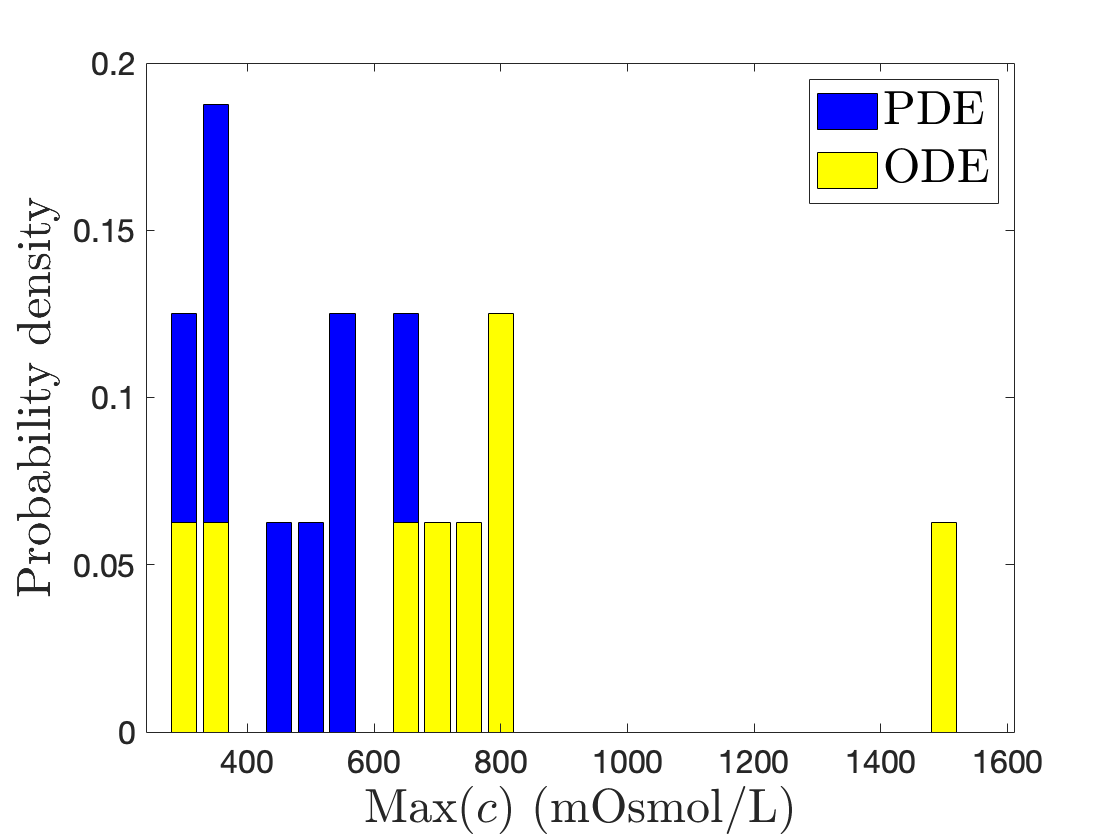}
\label{fig:c_h_a}}
\subfloat[][Minimum TF thickness]{\includegraphics[scale=.16]{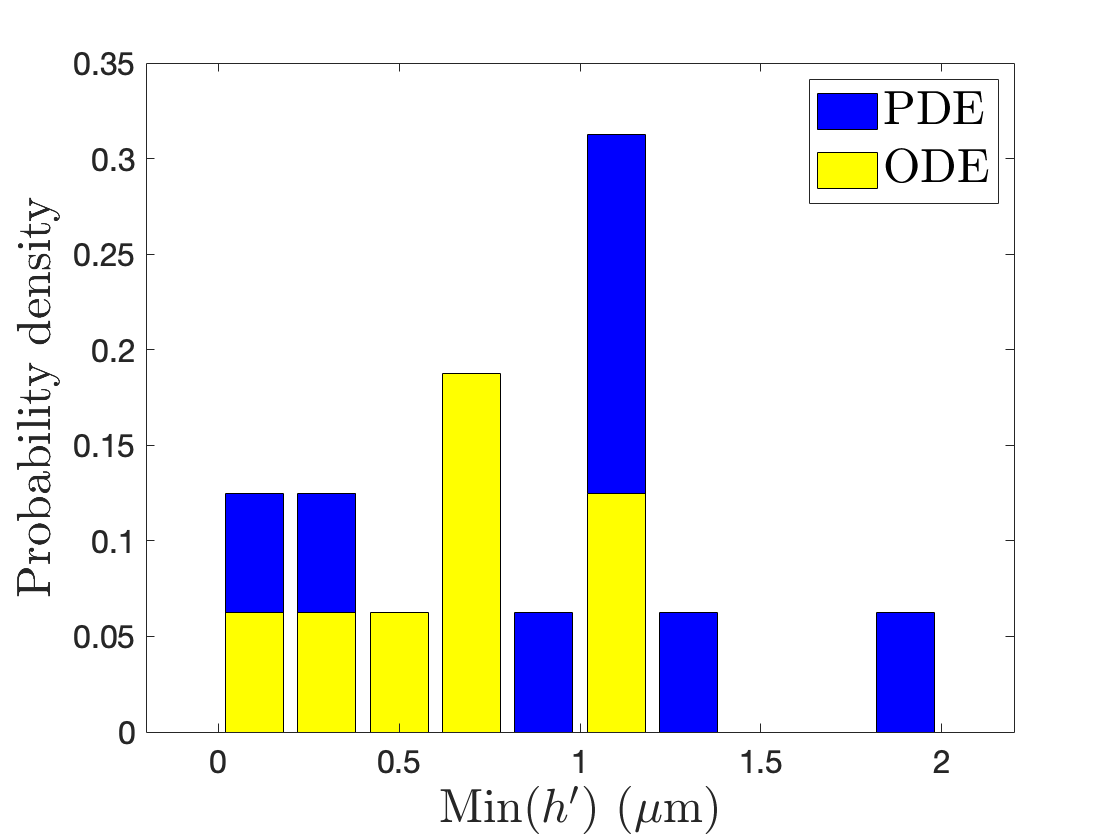}\label{fig:c_h_b}}
\caption{\footnotesize{Histograms of maximum osmolarity and minimum TF thickness (final times of fit).}
}
\label{fig:c_h}
\end{figure}

In Figure \ref{fig:c_h_a} we compare the maximum osmolarity values of the PDE and best fit ODE models. A bin size of 50 mOsmol/L was used. Both PDE and ODE peak osmolarity estimates are reasonable compared to other experimental and theoretical values\cite{liu09,peng2014}. The ODE results show greater variation and exhibit larger maximal values on average. Peng \textit{et al.}\cite{peng2014b} and Braun \textit{et al.}\cite{braun2018} showed that diffusion reduces the peak osmolarity in TBU, which is only relevant for a model with spatial variation. As such, the lower PDE maximum osmolarity values are expected. Further, Braun \textit{et al.}\cite{braun2015} reported theoretical osmolarity values up to ten times the isotonic level, and so our largest osmolarity value, which is just over five times isotonic, is not unreasonable. 

Figure \ref{fig:c_h_b} records the minimum thicknesses of the PDE and best ODE fits. A bin size of 0.2 $\mu$m was used. There is significant overlap of the PDE and ODE model results, suggesting that the simplified version can capture the end behavior of TF dynamics with a sufficient level of accuracy. The minimum TF thickness values are larger on average for the PDE models; this may be explained by the lag of the theoretical FL intensity behind the experimental data at later times in many of the PDE fits\cite{luke2020,luke2021}.

 Overall, there is more variability in the ODE results than the PDE results. We may overfit the subtleties of the dynamics with four parameters in the case (M) model, especially when the few data points of the central dynamics are essentially linear. Further, in some instances, the dynamics of the case (D) and (M) models are nearly indistinguishable, suggesting the additional parameter in the (M) model that mimics steady outward flow may not be necessary. The PDE data, which combines spatial and temporal information, is only fit with three parameters, reducing the likelihood of overfitting. Slight differences in the experimental data that are likely due to noise can affect the optimal parameters. Better time resolution would help get rid of the influence of outlier time levels on the optimization.

In order to compare with the PDE results, we scale with a characteristic length and horizontal velocity whose meanings are less clear in the context of the ODE model. The time scale we use is more of a characteristic time to bend the curve rather than the time to an overall decrease in FL intensity.
 We show the PDE results only for comparison; this ODE fitting process can be used on many other instances.

\section{Conclusions and future perspectives}
\label{sec:conc}

We fit the same data as in Luke \textit{et al.}\cite{luke2021} with simplified models to validate our ODE fitting procedure and find good qualitative agreement of PDE and ODE results in most instances. The ODE fitting procedure provides a relatively quick process that returns important information about the TBU instance, including parameters that cannot currently be measured directly \textit{in vivo}.

 We are working on a machine learning approach to automatically identify breakup instances and fit the central data with our ODE models. This strategy could be applied on a large scale to obtain statistical information about a wide range of TBU shapes.

%

\section*{Acknowledgements}
This work was supported by National Science Foundation grant DMS 1909846. The content is solely the responsibility of the authors and does not necessarily represent the official views of the funding sources.

\appendix

\section{Appendix}
\label{sec:app}

\subsection{Governing dimensional equations}

We derive the case (M) model. We use Cartesian coordinates $(x',z')$ to denote the position and $\vec{u}' = (u',v')$ to denote the fluid velocity. Primes denote dimensional quantities. The TF is modeled as an incompressible Newtonian fluid on $0 < x' < X_0$ and $0 < z' < h'(x',t')$, where $h'(x',t')$ denotes the thickness of the film. Conservation of mass of TF fluid is given by
\begin{equation}
\nabla' \cdot \vec{u}' = 0.
\end{equation}
At the film/cornea interface $z' = 0$, we require osmosis across a perfect semipermeable membrane:
\begin{equation}
u' = 0, \quad v' = P_o V_w (c'-c_0),
\end{equation}
where $c'$ is the osmolarity. The membrane permeability is given by $P_o$, the molar volume of water is $V_w$, and $c_0$ is the isotonic osmolarity. The kinematic condition at the fluid/air interface is given by \begin{equation}
\partial_{t'} h' = v'|_{z' = h'} - u'|_{z' = h'} \partial_{x'}h' - J',
\end{equation} where $J'$ is evaporation. Since we have assumed the film is spatially uniform, we have $\partial_{x'} h' = 0$, and thus
\begin{equation} \dot{h'} = v'_{z' = h'}-J'.
\end{equation}
The dot indicates an ordinary derivative in time. We assume a combination of constant and decaying extensional flow:
\begin{equation}
u' = (a' + b_1'e^{-b_2't'})x',
\label{eq:u}
\end{equation}
where $a'$ is a constant flow rate, and $b_1'$ and $b_2'$ are a flow and decay rate, respectively.

\subsection{Scalings}
\label{sec:scale}

The governing equations can be nondimensionalized using the following scalings:
\begin{equation}
c' = c_0 c, \quad \quad
h' = dh,
\quad \quad
t' = (\ell/U)t,
\quad \quad
u' = Uu,
\quad \quad
f' = f_{\text{cr}} f,
\quad \quad
J' = \rho U J.
\label{eq:scalings1}
\end{equation}
The following scalings are a result of the nondimensionalization process:
\begin{equation}
v' = \epsilon U v,
\quad \quad
a' = (U/\ell) a,
\quad \quad
b_1' = (U/\ell) b_1,
\quad \quad
b_2' = (U/\ell) b_2.
\label{eq:scalings2}
\end{equation}
Dimensional quantities are denoted by primes. We follow the same scalings used in Luke \textit{et al.}\cite{luke2021} to compare with their results.  We take the length of the trial as $t_s$ and then $U = \ell/t_s$, where
\begin{equation}
\ell = \left( \frac{t_s \sigma_0 d^3}{\mu} \right)^{1/4}.
\end{equation}
Dimensional parameters used in the model are summarized in Table \ref{table:dim}.

\begin{table}
\centering
\begin{tabularx}{\linewidth}{|p{.12\linewidth}|p{.27\linewidth}|p{0.28\linewidth}|X|}
\hline
\multicolumn{4}{|c|}{\cellcolor{jmotablecolor} \textbf{Dimensional parameters}}                                                                                                                           \\ \hline
\cellcolor{jmotablecolor} Parameter       & \cellcolor{jmotablecolor} Description                      & \cellcolor{jmotablecolor} Value                                                & \cellcolor{jmotablecolor} Reference                                                \\ \hline
$\rho$          & Density                          & $10^3$ kg $\cdot$ m$^{-3}$                           & Water                                                    \\
$d$             & Initial TF thickness             & $2-5 \times 10^{-6}$ m                               & Calculated\cite{luke2021} \\
$f_0'$ & Init.\ FL concentration & $0.259-0.4$ \% & Calculated\cite{luke2021} \\
$v'$      & Evaporative thinning rate               & $0.5 - 25$ $\mu$m/min                                      & Nichols \textit{et al.}\cite{nichols2005} \\
$a'$ & \begin{tabular}[l]{@{}c@{}}Constant extensional \\flow rate\end{tabular} & $-0.201 - 1.74$ $s^{-1}$ & Calculated \\
$b_1'$ & \begin{tabular}[l]{@{}c@{}}Decaying extensional \\flow rate\end{tabular} & $-0.564 - 3.69$ $s^{-1}$ & Calculated \\
$b_2'$ & Decay rate & $0.0421 - 12.8$ $s^{-1}$ & Calculated \\
$V_w$           & Molar volume of water            & 1.8 $\times 10^{-5}$ m$^3 \cdot $ mol$^{-1}$         & Water                                                    \\
$c_0$ & Isotonic osmolarity & $300$ mOsmol/L &  Lemp \textit{et al.}\cite{lemp2011} \\
$P_o$           & Permeability of cornea           & $12.1 \times 10^{-6}$ m/s                            & Braun \textit{et al.}\cite{braun2015} \\
$\epsilon_f$    & \begin{tabular}[l]{@{}c@{}}Naperian extinction \\coefficient\end{tabular}  & $1.75 \times 10^7$ m$^{-1}$ M$^{-1}$                  & Mota \textit{et al.}\cite{mota1991} \\
$\mu$ & Viscosity & $1.3$ $\times 10^{-3} \ \text{Pa} \cdot \text{s}$ & Tiffany\cite{tiffany1991} \\
$\sigma_0$ & Surface tension & $0.045$ N $\cdot$ m$^{-1}$ & Nagyog\'{a} \& Tiffany\cite{nagyova1999} \\
$\ell$ & Characteristic length & $0.138 - 0.412$ mm & Calculated \\
$U$ & Characteristic velocity & $0.0560 - 0.0990$ mm/s & Calculated \\
$t_s$ & Time scale & $1.75 - 6.6$ s & Fit interval\cite{luke2021} \\ \hline
\end{tabularx}
\caption{The dimensional parameters used are shown. The range of estimates for thinning rates are from point measurements; this range includes the values given by our optimization.}
\label{table:dim}
\end{table}

\begin{table}
\centering
\begin{tabularx}{\linewidth}{|X|p{0.33\linewidth}|X|X|}
\hline
 \multicolumn{4}{|c|}{ \cellcolor{jmotablecolor} \textbf{Non-dimensional parameters with typical values}} \\ \hline
\cellcolor{jmotablecolor} Parameter & \cellcolor{jmotablecolor} Description   & \cellcolor{jmotablecolor} Expression                                  & \cellcolor{jmotablecolor} Value  \\  \hline
$\epsilon$ & Aspect ratio & $d/\ell$ & 0.0130 \\
$P_c$   & Permeability of cornea    & $\displaystyle P_o V_w c_0/(\epsilon U)$              & 0.0653  \\
$\phi$   & Nondimensional Napierian extinction coefficient   & $\displaystyle \epsilon_f f_{\text{cr}} d$                & 0.279  \\ \hline
\end{tabularx}
\caption{Dimensionless parameters that arise from scaling the dimensional fluid mechanics problem. The values given are based upon the values of Table \ref{table:dim} and those used to generate Figure \ref{fig:plot_h}.}
\label{table:nondim}
\end{table}

FL concentration is typically reported as a percentage in the ocular literature. For a particular FL concentration $f'$ given as a percentage, this quantity is converted to molar as $f_M'$ by

\begin{equation}
f_M' = \frac{\rho}{M_w}\frac{f'}{100},
\end{equation}
where $\rho$ is the density of water (Table \ref{table:dim}) and $M_w$ is the molecular weight of sodium fluorescein (approximately 376 g/mol). Critical FL concentration $f_{\text{cr}}$, 0.2\%, makes an 0.0053 M solution when dissolved in water. This conversion of $f_{\text{cr}}$ to molar is necessary to compute the dimensionless Napierian extinction coefficient $\phi$ (Table \ref{table:nondim}).

\subsection{Derivation of TF equations}
\label{sec:app_deriv}

Using the scalings \ref{eq:scalings1}, \ref{eq:scalings2}, we nondimensionalize the governing equations.
From the nondimensional version of Equation \ref{eq:u}, we have that $u_x = a + b_1 e^{-b_2 t}.$
In Cartesian coordinates, conservation of mass is given by $u_x + v_z = 0$.  Integrating this equation over the vertical domain, we have
$$ 0 = \int_0^h \left[\partial_x u + \partial_z v \right] dz = -(a + b_1  e^{-b_2 t})h + \partial_t h + J - P_c(c-1),$$
where we've used the independence of $u_x$ from $z$. Rewriting this result as a differential equation for $h$, we have
\begin{equation}
\dot{h} = -(a + b_1  e^{-b_2 t})h - J + P_c(c-1).
\label{ODE_h}
\end{equation}
Our transport equation for $c$ is (without any spatial terms):
\begin{equation}
h \dot{c} = Jc - P_c(c-1)c.
\label{ODE_c}
\end{equation}
Multiplying Eq. (\ref{ODE_h}) by $c$ and adding the result to Eq. (\ref{ODE_c}), we have an ODE for the product $hc$:
$$ \dot{(hc)} = -(a + b_1  e^{-b_2 t}) hc.$$
Separating and integrating gives
\begin{equation}
hc = A \exp\left(-a t + \frac{b_1}{b_2} e^{-bt} \right),
\end{equation}
where $A$ is an arbitrary constant of integration. Using $h(0) = c(0) = 1$ and $f(0) = f_0$ nondimensionally, we solve for $c$ and $f$:
\begin{equation}
c = \frac{1}{h} \exp\left[-a t + \frac{b_1}{b_2} \left(e^{-b_2t} - 1 \right) \right], \quad f = \frac{f_0}{h} \exp\left[-a t + \frac{b_1}{b_2} \left( e^{-b_2 t} - 1 \right) \right].
\end{equation}
Replacement in our ODE for $h$ gives the equations shown in the text.

The equation for fluorescent intensity is
\begin{equation}
I = I_0 \frac{1 - e^{-\epsilon_f h' f'}}{\left(f'_{\text{cr}}\right)^2 + (f')^2},
\label{Idim}
\end{equation}
where $\epsilon_f$ is the molar extinction coefficient, $f_{\text{cr}}'$ is the critical fluorescein concentration, and $I_0$ is a normalization factor.
The nondimensional version with $f$ eliminated is given in the text.

%
%
%

\typeout{get arXiv to do 4 passes: Label(s) may have changed. Rerun}

\end{document}